\documentclass[twocolumn]{aastex63}
\received{---}
\revised{---}
\accepted{---}
\submitjournal{ApJ}

\shorttitle{Binaries as Density Probes}
\shortauthors{Rose et al.}
\graphicspath{{./}{figures/}}

\begin{document}

\title{On Socially Distant Neighbors: \\ Using Binaries to Constrain the Density of Objects in the Galactic Center}


\correspondingauthor{Sanaea C. Rose}
\email{srose@astro.ucla.edu}

\author{Sanaea C. Rose}
\affiliation{Department of Physics and Astronomy, University of California, Los Angeles, CA 90095, USA}
\affiliation{Mani L. Bhaumik Institute for Theoretical Physics,
University of California, Los Angeles,
CA 90095, USA}

\author{Smadar Naoz}
\affiliation{Department of Physics and Astronomy, University of California, Los Angeles, CA 90095, USA}
\affiliation{Mani L. Bhaumik Institute for Theoretical Physics,
University of California, Los Angeles,
CA 90095, USA}

\author{Abhimat K. Gautam}
\affiliation{Department of Physics and Astronomy, University of California, Los Angeles, CA 90095, USA}

\author{Andrea M. Ghez}
\affiliation{Department of Physics and Astronomy, University of California, Los Angeles, CA 90095, USA}

\author{Tuan Do}
\affiliation{Department of Physics and Astronomy, University of California, Los Angeles, CA 90095, USA}

\author{Devin Chu}
\affiliation{Department of Physics and Astronomy, University of California, Los Angeles, CA 90095, USA}



\author{Eric Becklin}
\affiliation{Department of Physics and Astronomy, University of California, Los Angeles, CA 90095, USA}




\begin{abstract}
Stars often reside in binary configurations. The nuclear star cluster surrounding the supermassive black hole (SMBH) in the Galactic Center (GC) is expected to include a binary population. In this dense environment, a binary frequently encounters and interacts with neighboring stars. These interactions vary from small perturbations to violent collisions. In the former case, weak gravitational interactions unbind a soft binary over the evaporation timescale, which depends on the binary properties as well as the density of surrounding objects and velocity dispersion. Similarly, collisions can also unbind a binary, and the collision rate depends on the density. Thus, the detection of a binary with known properties can constrain the density profile in the GC with implications for the number of compact objects, which are otherwise challenging to detect. We estimate the density necessary to unbind a binary within its lifetime for an orbit of arbitrary eccentricity about the SMBH. We find that the eccentricity has a minimal impact on the density constraint. In this proof-of-concept, we demonstrate that this procedure can probe the density in the GC using hypothetical young and old binaries as examples. Similarly, a known density profile provides constraints on the binary orbital separation. Our results highlight the need to consider multiple dynamical processes in tandem. In certain cases, often closer to the SMBH, the collision timescale rather than the evaporation timescale gives the more stringent density constraint, while other binaries farther from the SMBH provide unreliable density constraints because they migrate inwards due to mass segregation.

\end{abstract}



\section{Introduction} \label{sec:intro}
Most galaxies have a $10^{6-9}$~M$_\odot$ supermassive black hole (SMBH) at their center \citep[e.g.,][]{Kormendy04,FerrareseFord05,KormendyHo13}. The Milky Way's center hosts the closest known SMBH, Sagittarius A*, surrounded by a dense nuclear star cluster. With a mass of $4 \times 10^6 \, M_\odot$, the SMBH dominates the gravitational potential in the inner parsec region \citep{Ghez+05,Gillessen+09}. The proximity of this environment presents a unique opportunity to broaden our understanding of the physical processes unfolding in galactic centers.


The nuclear star cluster is largely dominated by a diffuse population of old ($ > 1$~Gyr) stars \citep{Nogueras-Lara+19,Schodel+20}, including several bright giants \citep[e.g.][]{Do+09}. 
Observations have also unveiled a population of approximately $4-6$ Myr-old stars within the central pc of the Galactic Center (GC) \citep[e.g.,][]{Lu+09,Bartko+10,Do+13a,Do+13b,FK+15}. A subset of those young stars, the so-called the S-star cluster, have eccentric orbits that are distributed isotropically within $\sim 0.04$ pc of the SMBH \citep{Ghez+05,Ghez+08,Schodel+03,Gillessen+09,Gillessen+17}, though a recent study suggests that the S-stars may be arranged in two discs \citep{Basel+20}. Additionally, observations have revealed a stellar disk structure in the inner parsec region \citep[e.g.,][]{Levin+03,Paumard+06,Lu+09,Bartko+09,Yelda+14}. 
Throughout this paper, when we refer to the GC, we are focusing on this inner region where the nuclear star cluster resides.
This unique environment, a stellar cluster embedded within the gravitational potential of a SMBH, is expected to yield several interesting phenomena such as hypervelocity stars \citep[e.g.,][]{Hills88,YuTremaine,GinsburgLoeb} and stellar binary mergers \citep[e.g.,][]{Antonini+10,Antonini+11,Prodan+15,Stephan+16,Stephan19}. These phenomena require the existence of binaries in this dense region.




Stars often reside in a binary configuration. Approximately $50\%$ of KGF stars and more than $70\%$ of OBA stars exist in binaries \citep[e.g.,][]{Raghavan+10}. However, few binaries have been observed in the GC. Approximately $0.05$ pc from the SMBH, the equal mass binary IRS 16SW has a period of $19.5$ days and total mass of approximately $100 \, M_\odot$ \citep{Ott+99,Martins06}. \citet{Pfuhl+14} confirm the existence of two additional binaries approximately $0.1$ pc from the SMBH: a short-period ($2.3$ days) eclipsing Wolf-Rayet binary and a long-period ($224$ days) low eccentricity ($e = 0.3$) binary. These confirmed binaries are the most direct evidence of the broader binary population expected to reside in the GC. A near-infrared variability study has detected $10$ binaries in the region of interest \citep{Dong+17a,Dong+17b}. Observational studies suggest that the OB binary fraction in the GC is comparable to that in young massive stellar clusters \citep[e.g.,][]{Ott+99,Rafelski+07}, and the eclipsing young OB binary fraction in the GC is consistent with the local fraction \citep{Gautam+19}. Furthermore, the theoretical study \citet{Stephan+16} suggests a $70\%$ binary fraction for the population from the most recent star formation episode, $6$ Myr ago, in the nuclear star cluster \citep[e.g.,][]{Lu+13}.

Few systems have been identified because the detection of binaries in the GC stellar population faces several observational challenges. These challenges include high extinction towards the GC and extreme stellar crowding near the supermassive black hole. Adaptive optics (AO) on large ground-based telescopes, allowing deep, diffraction-limited observations of the GC stellar populations, can overcome some of these limitations. However, AO imaging limits photometric precision \citep[e.g.,][]{Schodel+10,Gautam+19}, while AO spectroscopic studies are not sensitive to fainter members of the GC stellar population \citep[e.g.,][]{Do+13a}. Furthermore, binary searches require large stellar sample sizes and frequent monitoring of the stellar population to measure sufficient photometric or spectroscopic variability. However, not many such surveys have yet been performed for the stellar population in the central half parsec of the GC.

However, other phenomena hint at the existence of binaries in the GC. The abundant X-ray sources there may trace to binaries in which a black hole accretes material from a stellar companion \citep{Muno+05,Cheng+18,Zhu+18,Hailey+18}, while \citet{Muno+06,Muno+09} and \citet{Heinke+08} attribute these X-ray sources to Cataclysmic Variables, a binary composed of a White Dwarf and main-sequence star. Hypervelocity stars likely originate from a binary that has been disrupted by the SMBH, ejecting one of the binary members from the GC \citep[e.g.,][]{Hills88, YuTremaine, GinsburgLoeb,Perets09a,Perets09}. More recently, a theoretical study, \citet{Naoz+18} explains puzzling disk properties with an abundance of binary systems.

In a dense environment like the GC, a binary system frequently encounters other stars. Several studies have explored the complex physics and implications of these encounters \citep[e.g.,][]{Heggie75,Hills75,HeggieHut93,RasioHeggie95,HeggieRasio96,BinneyTremaine,Hopman09,Leigh+16,Leigh+18,HamersSamsing19a}.
If a passing star approaches the binary with impact parameter on the order of the binary separation, it interacts more strongly with the closer binary member. This interaction imparts energy to the binary system and causes the binary to widen. Over a long period of time, many such encounters eventually unbind the binary. The evaporation timescale refers to the amount of time necessary for the binary to undergo this process \citep[see the derivation in][]{BinneyTremaine}. This timescale depends on the binary's characteristics as well as environmental properties such as the stellar number density and velocity dispersion. Additionally, stars in the binary can collide with passing objects with a timescale that also depends on the density of the system's environment \citep[e.g.,][]{Sigurdsson&Phinney93,Fregeau+04,BinneyTremaine}. The survival of the binary over its lifetime therefore depends on the surrounding density; too dense an environment results in the binary's destruction. Given these relations, a binary system with a known age provides an upper limit on the local density and can constrain the density profile in the GC.

Mass segregation is expected to concentrate the dense remnants of massive stars in the central pc of the GC \citep[e.g.,][]{Shapiro+78,Cohn+78,Morris93,MiraldaEGould00,Baumgardt+04}. Numerous studies explore the expected number of stellar mass black holes and their influence on the density profile of the GC \citep[e.g.,][]{MiraldaEGould00,Freitag+06,AlexanderHopman+09,Merritt10,Aharon+16}. In particular, if the GC evolved in isolation, it should be dynamical relaxed \citep[e.g.,][]{Bar-Or+13}, resulting in a simple, cusp-like power law density profile \citep[e.g.,][]{BahcallWolf76,AlexanderHopman+09,Keshet+09,Aharon+16}\footnote{Certain physical processes may also modify the density profile. For example, binary disruption by the SMBH can steepen the density cusp \citep{FragioneSari18}}. However, observations indicate that the profile may be shallower \citep[e.g.,][]{Buchholz+09,Do+09,Bartko+10,Gallego+18,Gallego+20,Schodel+14,Schodel+18,Schodel+20}. A density constraint can clarify the profile and the number and distribution of stellar remnants, which are difficult to detect.

We expand upon the framework presented in \citet{AlexanderPfuhl14}, the first use of a binary to constrain the density of the GC. In particular, we derive the evaporation timescale for a binary with an arbitrary eccentricity about the SMBH. We consider several dynamical processes, such as collisions and two body relaxation, in this proof-of-concept of binaries as probes of the GC density. Similarly, a known density profile can be used to infer a binary's orbital configuration. We begin by outlining the equations that describe relevant dynamical processes, including evaporation, two-body relaxation, the Eccentric Kozai-Lidov mechanism, and collisions, in Section~\ref{sec:maths}. Section~\ref{sec:outcome} summarizes the qualitative outcomes of binary systems subjected to competing dynamical processes. We explore the use of binaries as a probe of the GC environment, specifically the underlying density distribution, and address the fates of older long-lived binaries in Section~\ref{sec:constraints}. Lastly, we constrain the parameter space for hypothetical systems by assuming that observed S-stars reside in a binary system in Section~\ref{sec:sstars}. We summarize our results in Section~\ref{sec:conclusion}.

\section{Equations}
\label{sec:maths} 
\subsection{The Evaporation Process} \label{sec:Evaporation}

The evaporation process describes the unbinding of a binary due to interactions with passing neighbors.\footnote{Binaries can also be ionized, that is, unbound in a single interaction with a neighbor. The associated timescale for this process is a factor of approximately $\ln \Lambda$ larger than the evaporation timescale \citep{Heggie75,BinneyTremaine}. We omit this process here because evaporation occurs on a faster timescale.} Derived using diffusion physics \citep{BinneyTremaine}, the evaporation timescale depends on several properties. Some of these parameters pertain to the binary itself, such as the initial semimajor axis, $a_{\rm bin}$ and mass $M_{\rm bin}$, while others describe the binary's environment. In the latter category, the density of the scatterers, likely stars, and their velocity dispersion affect this timescale.

The evaporation timescale is inversely proportional to the density. The higher the density of scatterers, which are stars and stellar remnants, the more frequent the encounters. We assume that the density of objects is spherically symmetric and write it as a function of $r_\bullet$, the distance from the SMBH:
\begin{eqnarray}
    \rho(r_\bullet) = \rho_0 \left( \frac{r_\bullet}{r_0}\right)^{-\alpha} \ , \label{eq:density}
\end{eqnarray}
where $\rho_0$ is the normalized density at $r_0$ and $\alpha$ is the slope of the power law.\footnote{We do not consider any broken power laws. However, in Section~\ref{sec:density}, the density constraint derived from a binary's survival depends very weakly on $\alpha$, suggesting that, despite our assumptions here, the procedure can be used even if a broken power law describes the density profile in the GC.} We use values $1.35 \times 10^6 \, M_\odot/{\rm pc}^3$ and $0.25 \, {\rm pc}$ for these parameters, respectively \citep{Genzel+10}. These values are very similar in magnitude to those presented in \citet{Schodel+18}.

Additionally, the evaporation timescale is proportional to the velocity dispersion since the relative speed between the binary and passing star determines the length of each encounter. For a low velocity dispersion, the passing star's gravitational force has longer to act on the binary, producing greater change in the binary's orbital configuration. The Jeans Equation relates the density distribution to the one dimensional velocity dispersion. We express the velocity dispersion as:
\begin{eqnarray}\label{eq:sigma}
    \sigma(r_\bullet) = \sqrt{ \frac{GM_{\bullet}}{r_\bullet(1+\alpha)}},
\end{eqnarray}
where $G$ is the gravitational constant, $\alpha$ is the slope of the density profile, and $M_{\bullet}$ denotes the mass of the SMBH \citep{Alexander99,AlexanderPfuhl14}.

We consider a binary at distance $r_\bullet$ from the SMBH. We adapt the evaporation timescale equation to depend explicitly on the distance $r_\bullet$ \citep{BinneyTremaine,AlexanderPfuhl14,Stephan+16}:
\begin{equation}\label{eq:tev}
    t_{ev} = \frac{\sqrt{3} \sigma(r_\bullet)}{32\sqrt{\pi} G \rho(r_\bullet) a_{\rm bin} \ln{\Lambda(r_\bullet)}} \frac{M_{\rm bin}}{\langle M_\ast \rangle}, \label{eq:mastereq}
\end{equation}
where $M_{\rm bin}$ is the total mass of the binary and $\langle M_\ast \rangle$ is the average mass of a star in the GC.\footnote{We assume that $n \langle M_\ast^2 \rangle \approx \rho \langle M_\ast \rangle$ from the evaporation timescale equation in \citet{AlexanderPfuhl14}. However, an alternative is defining $M_\ast = \langle M_\ast^2 \rangle / \langle M_\ast \rangle$ such that $\rho M_\ast$ appears in the denominator of Eq.~\ref{eq:mastereq} following the notation of \citet{KocsisTremaine11}.} The evaporation timescale also depends on the Coulomb logarithm $\ln \Lambda$, where $\Lambda$ is the ratio of the maximum to minimum impact parameter. In the evaporation process, $b_{max}= a_{\rm bin}/2$ for the passing star to interact more strongly with one of the binary members. Otherwise, the encounter will affect the center of mass. The strongest deflection, $90^{\circ}$, gives $b_{min}$. We obtain the expression from \citet{AlexanderPfuhl14}:
\begin{equation}
     \Lambda = 2 \frac{\sigma^2}{v_{\rm bin}^2} =\frac{2 a_{\rm bin} M_{\bullet}}{M_{\rm bin}r_\bullet(1+\alpha)} \ .
\end{equation}
 In the last transition, we substitute Eq.~(\ref{eq:sigma}) for $\sigma$ and $GM_{\rm bin}/a_{\rm bin}$ for the orbital velocity of the binary, averaged over the mean anomaly. This approximation assumes that the inner binary orbital timescale is shorter than the orbital timescale about the SMBH.

\begin{figure}
	\includegraphics[width=\columnwidth]{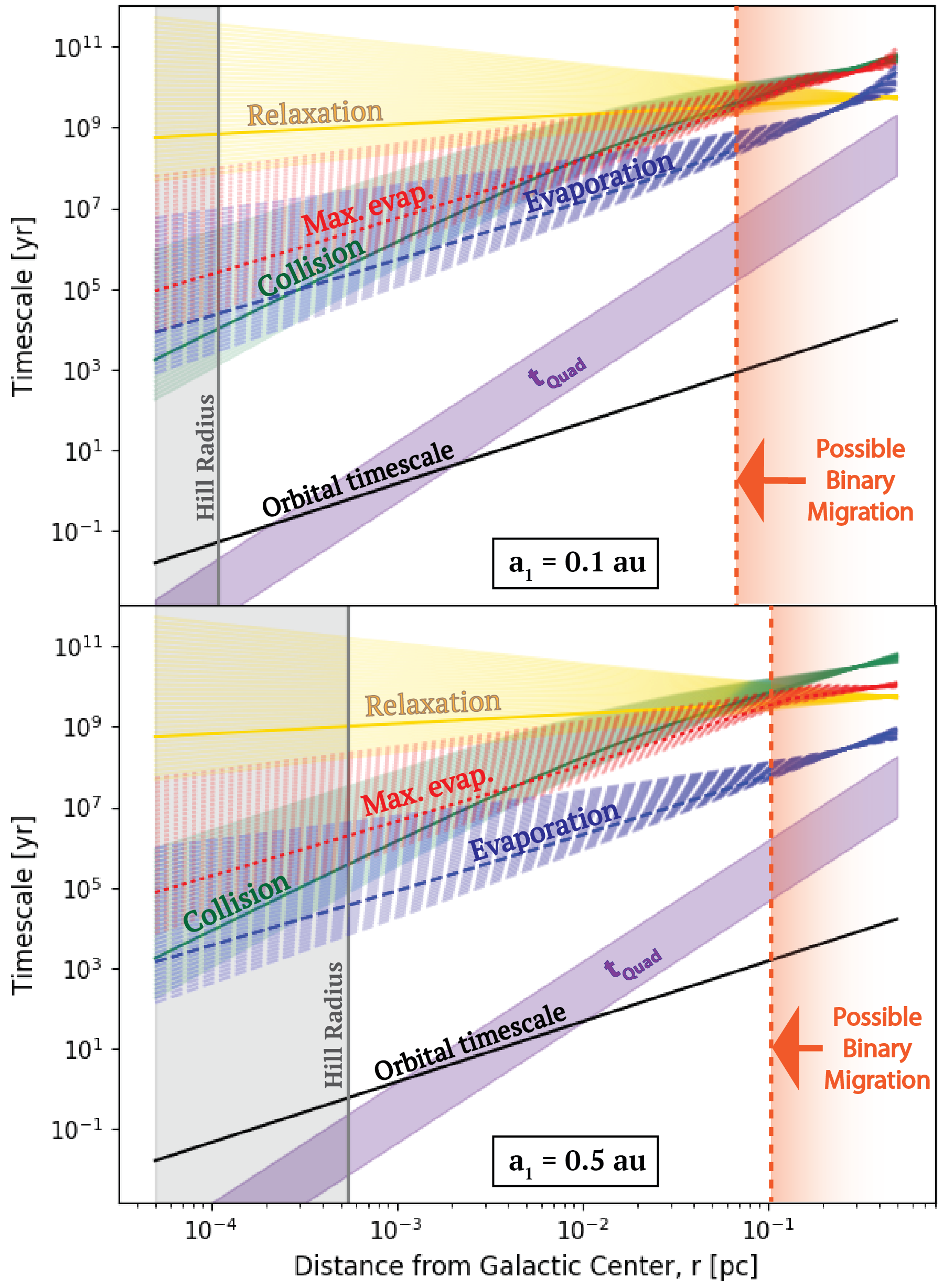}
    \caption{Two examples for the relevant timescales in the problem, as a function of the distance of the binary from the SMBH. We consider an equal mass binary with $M_{\rm bin} = 2 M_\odot$ and semimajor axis $0.1$~au and $0.5$~au for the top and bottom panels, respectively. For the timescales that depend on the density, we consider a range of power laws from  $\alpha = 1$ to $\alpha = 2$. All of the timescales increase with decreasing $\alpha$. Relevant timescales include the evaporation timescale from Equation (\ref{eq:tev}) (dark blue), the evaporation timescale with the history parameter (red, labeled Max. Evap.), the relaxation timescale (gold), the collision timescale (green), and the EKL quadrupole timescale (purple). The darkest lines have $\alpha = 1.75$ \citep{BahcallWolf76}.}
    \label{fig:timescales}
\end{figure}

Combining these equations and assumptions, we find that the evaporation time has $r_\bullet$ dependence:
\begin{equation}
    t_{ev} \propto \frac{r_\bullet^{\alpha-1/2}}{\ln{\left(\beta/r_\bullet\right)}} \ ,
\end{equation}
where
\begin{equation}
    \beta =\frac{2 a_{\rm bin} M_{\bullet}}{M_{\rm bin}(1+\alpha)} \ .
\end{equation}
We illustrate the dependency of the evaporation timescale as a function of distance in Figure \ref{fig:timescales}. Specifically, we plot the evaporation time of an equal mass $M_{\rm bin} = 2 \, M_\odot$ binary with $0.1$ and $0.5$~au semimajor axis for $\alpha = 1$ to $\alpha = 2$ in Figure \ref{fig:timescales} in dark blue. The darkest line represents $\alpha = 1.75$, the profile for a dynamically relaxed single mass system \citep{BahcallWolf76}. As the evaporation process relies on weak encounters, the true evaporation time of a binary system likely does not differ substantially from the evaporation timescale \citep{Perets+07}. We extend the axes in Figure~\ref{fig:timescales} to extreme values close to the SMBH, where our assumptions may break down, in particular regarding a continuous distribution of objects. However, we note that a recent analysis of S0-2 observations suggests that low-mass objects may still reside interior to its orbit \citep[e.g.,][]{Naoz+20}.

\subsubsection{Binaries Soften with Time}
The evaporation process requires that a binary begins in a soft configuration. A soft binary has a gravitational binding energy that is less than the kinetic energy of the neighboring stars:
\begin{eqnarray} \label{eq:softness}
     s = \frac{E}{\langle M_{\ast} \rangle \sigma^2} < 1,
\end{eqnarray}
where $E = G m_1 m_2/(2a_{\rm bin})$ and $\langle M_{\ast} \rangle$ is the average stellar mass in the GC. This configuration allows a passing star to interact more strongly with one of the binary members. This condition places a minimum on the semimajor axis a binary can have to evaporate. Following \citet{AlexanderPfuhl14}, we refer to $s$ as the softness parameter.

Additionally, soft binaries tend to soften over time. The evaporation timescale depends on $a_{\rm bin}$. However, as $a_{\rm bin}$ increases with time, the evaporation timescale depends on when in its lifetime the binary is observed. The birth configuration, namely the initial $a_{\rm bin}$, should determine the true evaporation timescale of the system. Assuming that the binary began harder and softened over its lifetime places an upper limit on the evaporation time:
\begin{eqnarray}\label{eq:tevmax}
     t_{ev,\mathrm{max}} &=& t_{ev} S_h \nonumber \\ &=& \frac{\sqrt{3} \sigma(r_\bullet)}{32\sqrt{\pi} G \rho(r_\bullet) a_{\rm bin} \ln{\Lambda(r_\bullet)}} \frac{M_{\rm bin}}{\langle M_\ast \rangle} S_h\ ,
\end{eqnarray}
where
\begin{equation}\label{eq:Sh}
S_h = \frac{s_{h}}{s_0} \ 
\end{equation} 
accounts for the binary's history, the widening the binary separation over time \citep{AlexanderPfuhl14}. We refer to $S_h$ as the history parameter. $s_0$ is the the softness parameter calculated at present time, when the binary is observed. $s_{h}$ finds the hardest possible initial configuration that the binary can have in order to place the most conservative overestimate on the evaporation time: 
\begin{equation}\label{eq:sh}
    s_{h} = \mathrm{min}[1,s(a_{\rm bin} = R_{1}+R_2)] \ ,
\end{equation} 
where $R_{1,2}$ are the initial zero-main sequence radii of the stars, estimating the binary as a contact binary at the beginning of it's life \citep{AlexanderPfuhl14}. In other words, $s_h$ chooses the tightest possible initial configuration to permit evaporation at birth, either a contact binary or at the limit for a soft binary, given by $s = 1$ (Eq.~(\ref{eq:softness})). 

Assuming the star's mass and the velocity dispersion have not changed over the binary's lifetime, the history parameter reduces to the ratio of the observed present day semimajor axis $a_{\rm bin}$ to the tightest possible initial semimajor axis $a_{\rm bin, min}$. Therefore, working from Eq.~(\ref{eq:tevmax}),
\begin{eqnarray}
 \label{eq:evmax_intuition}
    t_{\rm ev, max} &=& t_{\rm ev}(a = a_{\rm bin}) \times \frac{a_{\rm bin}}{a_{\rm bin, min}} \nonumber \\ &\approx& t_{\rm ev}(a = a_{\rm bin,min}) \ .
\end{eqnarray} 
If we assume no mass loss due to stellar evolution and neglect any changes in the Coulomb logarithm, scaling by the history parameter is equivalent to simply calculating the evaporation timescale using the tightest possible configuration $a_{\rm bin, min}$.

We include a conservative maximum evaporation time (red) in Figure \ref{fig:timescales} by scaling the dark blue curves by factor $S_h$. The true evaporation time lies between these two curves.

\subsubsection{Orbit Averaging}
\label{sec:ecc_avg}
In the GC, the density and velocity dispersion depend on the distance from the SMBH, around which any binaries must also orbit. Therefore, the evaporation timescale also depends on the eccentricity of the binary's orbit about the SMBH. A binary on an eccentric orbit will pass through the denser, more energetic inner regions of the GC unlike a binary on a circular orbit with the same semimajor axis. To account for the eccentricity's effect, we average the contributions of each segment of the orbit. We weight each segment by the amount of time the binary spends there.

We consider a stellar binary with orbital parameters $a_{\rm bin},e_{\rm bin}$ on an eccentric orbit about the Galactic Center. We refer to the stellar binary as the inner orbit. We define the outer orbit, that of the binary about the SMBH, to have orbital parameters $a_\bullet,e_\bullet$. Averaging over the outer orbit, we obtain the evaporation time as a function of orbital parameters $a_\bullet$ and $e_\bullet$. Specifically, we average over the canonical coordinate $M_\bullet$, the mean anomaly of the outer orbit:
\begin{eqnarray} \label{eq:meanAnomAvg}
   t_{ev} \propto \frac{1}{2\pi} \int_0^{2\pi} \frac{r_\bullet^{\alpha-1/2}}{\ln{\left(\beta/r_\bullet\right)}} dM_\bullet.
\end{eqnarray}

However, to integrate, we make a coordinate transformation from the mean anomaly to the eccentric anomaly, $E_\bullet$, using
\begin{eqnarray}
    r_\bullet = a_\bullet(1-e_\bullet \cos{E_\bullet}) \label{eq:r2}
\end{eqnarray}
and $dM_\bullet = r_\bullet/a_\bullet dE_\bullet$.

We assume that the Coulomb Logarithm does not depend on $r_\bullet$, i.e. the logarithm changes very slowly and can be considered constant in the integral. We arrive at the integral:
\begin{eqnarray}
   t_{\rm ev} \propto \frac{1}{2\pi} \int_0^{2\pi} a_\bullet^{\alpha-1/2}\left(1-e_\bullet \cos{E_\bullet}\right)^{\alpha+1/2} dE_\bullet
\end{eqnarray}
with the result $t_{\rm ev} \propto a_\bullet^{\alpha-1/2} \times f(e)$
where
\begin{eqnarray}
    f(e_\bullet) &=& \frac{(1-e_\bullet)^{\alpha+\frac{1}{2}}}{2} {}_2 F_1 \left(\frac{1}{2},-\frac{1}{2}-\alpha;1;\frac{2e_\bullet}{e_\bullet-1}\right) \nonumber \\& +& \frac{(1+e_\bullet)^{\alpha+\frac{1}{2}}}{2} {}_2 F_1 \left(\frac{1}{2},-\frac{1}{2}-\alpha;1;\frac{2e_\bullet}{e_\bullet+1}\right) \ 
\end{eqnarray}
and ${}_2 F_1$ is the hyper geometric function. 
Including the other parameters, we find that
\begin{eqnarray}
    t_{\rm ev} = \frac{\sqrt{3} \sigma(a_\bullet)}{32\sqrt{\pi} G \rho(a_\bullet) a_{\rm bin} \ln{\Lambda(a_\bullet)}} \frac{M_{\rm bin}}{\langle M_\ast \rangle} \times f(e_\bullet) \ , \label{eq:mastereq}
\end{eqnarray}
where $\sigma$, $\rho$, and the Coulomb logarithm are all evaluated at $a_\bullet$. In Section~\ref{sec:density}, we rearrange this equation to obtain the maximum density at a distance $a_\bullet$ from the SMBH. Note that $f(e)$ is always of order unity, indicating that the eccentricity of an orbit about the SMBH makes little difference to the evaporation time. We remind the reader that the maximum evaporation timescale is achieved by multiplying Eq.~(\ref{eq:mastereq}) by the history parameter, $S_h$ (see Eq.~(\ref{eq:tevmax})).

\begin{figure}
	\includegraphics[width=\columnwidth]{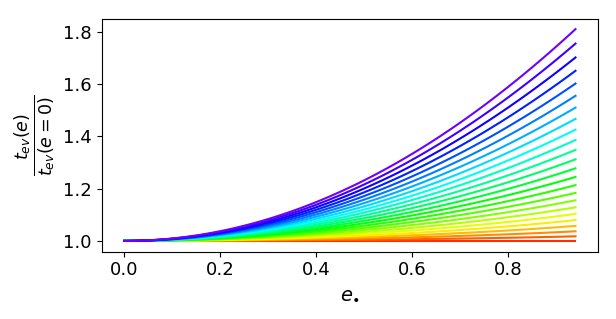}
    \caption{The evaporation time for a binary on an eccentric orbit with semimajor axis $1$ au normalized by that of a circular orbit. We vary the density distribution of stars by assuming different values of $\alpha$, from $\alpha = 0.5$ (red), which gives constant evaporation time, to $\alpha = 2$ (purple).}
    \label{fig:eccentricity}
\end{figure}

In Figure \ref{fig:eccentricity}, we plot the evaporation timescale as a function of eccentricity normalized by that of a circular orbit of the same semimajor axis. We vary the stellar density as a function of radius by using values of $\alpha$ from $0.5$ to $2$. Binaries on eccentric orbits have a longer evaporation timescale, and this effect becomes more pronounced with increasing $\alpha$. This result indicates that the evaporation timescale for a binary on an eccentric outer orbit will have the same order of magnitude as the circular case. This minimal change stems from the fact that the binary will spend more time near the apoapsis of its orbit, offsetting the effects of reaching denser, more energetic regions closer to the SMBH. The environmental conditions at the distance $a_\bullet$ therefore give a good approximation of the evaporation time.

\subsection{Two-Body Relaxation}

The evaporation process requires that the maximum impact parameter must be about $ a_{\rm bin}/2$ to ensure that the passerby interacts more strongly with one of the binary members as opposed to the binary center of mass. However, the interactions that fail to meet this criterion still impact the dynamical evolution of the binary. Interactions acting on the center of mass change the binary's overall trajectory. Over a relaxation time, these interactions alter the outer orbit.

For a single mass system, the two-body relaxation timescale can be written as: 
\begin{eqnarray} \label{eq:t_rlx}
t_{\rm relax} = 0.34 \frac{\sigma^3}{G^2 \rho \langle M_\ast \rangle \ln \Lambda_{\rm rlx}}
\end{eqnarray}
\citep[][Eq.~(7.106)]{Spitzer1987,BinneyTremaine}.
We use this equation to gauge the timescale over which the orbit of the binary about the SMBH changes as $m_{\rm bin}\sim \langle M_\ast \rangle$ in this case. Additionally, the timescale over which the outer orbit changes through binary interactions with single stars is similar to the relaxation timescale \citep{Hopman09}. Interactions at all impact parameters perturb the binary's trajectory about the SMBH, so the Coulomb logarithm in the relaxation timescale $\ln \Lambda_{\rm rlx}$ differs from that in the evaporation timescale. The strongest deflection still gives $b_{\rm min}$. However, $b_{\rm max}$ becomes $r_\bullet$ to encompass all interactions from objects interior to the binary's orbit about the SMBH. In Figure~\ref{fig:timescales}, we plot the relaxation timescale for a variety of $\alpha$ in gold with the dark line representing $\alpha = 1.75$.

Perhaps more appropriate for binary systems is the mass segregation timescale, which also derives from the relaxation process and has a similar form to Eq.~(\ref{eq:t_rlx}) \citep[e.g.,][]{BonnellDavies98,Spitzer1987,Merritt06}. Binary systems sink inwards according to their masses, which are higher on average than that of the surrounding objects \citep[e.g.,][]{MathieuLatham86,Geller+13,Antonini14}. The binary migrates towards the SMBH on the mass segregation timescale $t_{\rm seg}\approx \langle M_\ast \rangle/m_{\rm bin} \times t_{\rm relax}$ \citep[e.g.,][]{Merritt06}. For more massive systems, we use this equation to calculate the timescale over which the outer orbit changes.\footnote{We focus on changes to the energy of the outer orbit, which affects the semimajor axis, through two-body relaxation. However, resonant relaxation processes also take place, altering the outer orbit's angular momentum magnitude and orientation \citep[e.g.,][]{RauchTremaine96,HopmanAlexander06,KocsisTremaine11}. Since the collision and evaporation timescales depend weakly on eccentricity (Figures 2 and 3), scalar resonant relaxation, which changes the angular momentum vector's magnitude, does not impact our procedure. Similarly, vector resonant relaxation, which changes the outer orbit's inclination, also does not affect our procedure, although it can help the system enter the EKL-favored regime \citep[e.g.,][]{Hamers+18}.}

\subsection{Inelastic Collisions}

In a dense environment like the GC, direct collisions between objects occur frequently. Accounting for gravitational focusing, the collision rate can be expressed as:
\begin{eqnarray} \label{eq:t_coll}
t_{\rm coll}^{-1} = 16 \sqrt{\pi} \frac{\rho(r_\bullet) \sigma(r_\bullet)}{\langle M_\ast \rangle} R_j^2 \left(1+\frac{v_{\rm esc}^2}{\sigma(r_\bullet)^2}\right),
\end{eqnarray}
where $v_{\rm esc} = \sqrt{2Gm_j/R_j}$ is the escape speed from one of the stars in the binary with mass (radius) $m_j$ ($R_j$) \citep[][Eq. (7.195)]{BinneyTremaine}. 
For a soft binary, we treat the collision separately from the evolution of the orbit. The primary will suffer a collision sooner than the less massive companion because of its enhanced cross-section. Therefore, we calculate the collision timescale using the primary star. We plot this timescale in green in Figure~\ref{fig:timescales}.

\begin{figure}
	\includegraphics[width=\columnwidth]{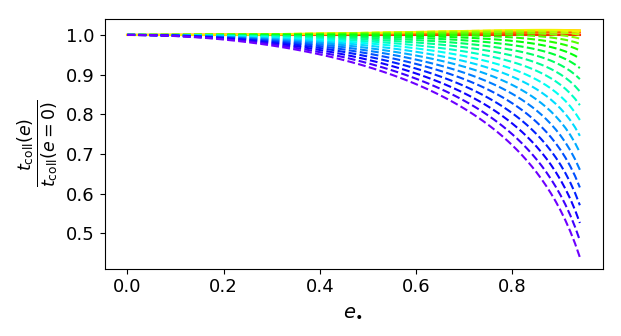}
    \caption{The collision timescale for a star on an eccentric orbit about the SMBH normalized by that of a circular orbit. We vary the density distribution of stars by assuming different values of $\alpha$, from $\alpha = 0.5$ (red) to $\alpha = 2$ (purple).}
    \label{fig:tcoll_vs_ecc.png}
\end{figure}

To find the collision timescale for a star on an eccentric orbit about the SMBH, we average the collision rate Eq.~(\ref{eq:t_coll}) over the mean anomaly similar to Eq.~(\ref{eq:meanAnomAvg}). After a coordinate transformation to the eccentric anomaly $E_\bullet$, over which we integrate, the collision rate for an eccentric orbit becomes
\begin{eqnarray} \label{eq:t_coll_ecc}
t_{\rm coll}^{-1} &=& 16 \sqrt{\pi} \frac{\rho(a_\bullet) \sigma(a_\bullet)}{\langle M_\ast \rangle} R_j^2 \nonumber \\ &\times& \left(f_1(e_\bullet)+f_2(e_\bullet)\frac{v_{\rm esc}^2}{\sigma(a_\bullet)^2}\right),
\end{eqnarray}
where
\begin{eqnarray}
    f_1(e_\bullet) &=& \frac{(1-e_\bullet)^{\frac{1}{2}-\alpha}}{2} {}_2 F_1 \left(\frac{1}{2},\alpha-\frac{1}{2};1;\frac{2e_\bullet}{e_\bullet-1}\right) \nonumber \\& +& \frac{(1+e_\bullet)^{\frac{1}{2}-\alpha}}{2} {}_2 F_1 \left(\frac{1}{2},\alpha-\frac{1}{2};1;\frac{2e_\bullet}{e_\bullet+1}\right) \
\end{eqnarray}
and
\begin{eqnarray}
    f_2(e_\bullet) &=& \frac{(1-e_\bullet)^{\frac{3}{2}-\alpha}}{2} {}_2 F_1 \left(\frac{1}{2},\alpha-\frac{3}{2};1;\frac{2e_\bullet}{e_\bullet-1}\right) \nonumber \\& +& \frac{(1+e_\bullet)^{\frac{3}{2}-\alpha}}{2} {}_2 F_1 \left(\frac{1}{2},\alpha-\frac{3}{2};1;\frac{2e_\bullet}{e_\bullet+1}\right)  .
\end{eqnarray}
Figure \ref{fig:tcoll_vs_ecc.png} shows the collision timescale as a function of eccentricity normalized by that of a circular orbit. The density has values of $\alpha$ ranging from $0.5$ (red) to $2$ (purple). Increasing the eccentricity increases the average collision rate, decreasing the collision timescale. The change becomes more pronounced for steeper density profiles but does not exceed a factor of two. As in Section~\ref{sec:ecc_avg}, we attribute this minimal change to the conditions near periapsis and apoapsis offsetting the other's impact.

Eq.~(\ref{eq:t_coll}) assumes that stars of equal mass compose the population, and its derivation involves averaging over the velocity distribution of the stars. In later sections, we violate this assumption by considering a massive star with less massive surrounding objects. We can estimate the collision timescale in this scenario using a simple $t_{\rm coll}^{-1} = n \sigma A$ calculation. The cross-section of interaction $A$ equals $\pi b^2$, where b is the maximum impact parameter for a physical collision. Accounting for gravitational focusing,
\begin{eqnarray}
     b^2 = r_c^2 + r_c \frac{2GM_t}{\sigma^2}\ ,
\end{eqnarray}
where $r_c$ is the sum of the radii of the interacting stars, $R_j + \langle R_\ast \rangle$, and $M_t = m_j + \langle M_\ast \rangle$. We estimate that the collision rate is
\begin{eqnarray} \label{eq:t_coll_main}
     t_{\rm coll}^{-1} = \frac{\pi \rho \sigma}{\langle M_\ast \rangle} \left(r_c^2 + r_c \frac{2GM_t}{\sigma^2}\right)\ .
\end{eqnarray}
In the limit that all stars are identical, we set $r_c = 2R_j$ to find that Eq.~(\ref{eq:t_coll}) is only a factor of two bigger than Eq.~(\ref{eq:t_coll_main}). This comparison suggests that averaging over the velocity distribution introduces a small numerical factor, increasing the collision rate. Averaging Eq.~(\ref{eq:t_coll_main}) over the eccentricity results in a very similar equation to Eq.~(\ref{eq:t_coll_ecc}):
\begin{eqnarray} \label{eq:t_coll_main_ecc}
     t_{\rm coll}^{-1} &=& \frac{\pi \rho(a_\bullet) \sigma(a_\bullet)}{\langle M_\ast \rangle} \nonumber \\ &\times& \left(f_1(e_\bullet)r_c^2 + f_2(e_\bullet)r_c \frac{2GM_t}{\sigma(a_\bullet)^2}\right)\ .
\end{eqnarray}


\subsection{Eccentric Kozai-Lidov Timescale}

Any binary system in the GC forms a triple system with the SMBH. Dynamical stability requires that this triple has a hierarchical configuration: the binary must have a much tighter (inner) orbit than that of its center of mass about the SMBH, referred to as the outer orbit. The condition $a_{\rm bin} << a_\bullet$ allows an expansion of the gravitational potential in terms of the small parameter $a_{\rm bin}/a_\bullet$ \citep{Kozai,Lidov}. The SMBH perturbs the binary's orbit, causing the eccentricity and inclination of the stellar binary to oscillate in the Eccentric Kozai-Lidov (EKL) Mechanism \citep[see the review][]{Naoz16}. The quadrupole term in the expansion has an associated timescale, which represents the timescale of the oscillations:
\begin{eqnarray}
t_{\rm EKL,quad} = \frac{16}{30\pi} \frac{M_{\rm bin}+M_\bullet}{M_\bullet} \frac{P_\bullet^2}{P_{\rm bin}} \left(1-e_\bullet^2\right)^{3/2},
\end{eqnarray}
where $P_{\rm bin}$ ($P_\bullet$) denotes the period of the inner (outer) orbit \citep[e.g.,][]{Antognini15,Naoz16}. We plot a range of quadrupole timescales in purple for $e_\bullet = 0$ to $e_\bullet = 0.95$ in Figure \ref{fig:timescales}.

These EKL oscillations can result in one of several outcomes. The oscillations can drive the eccentricity to extreme values and cause the inner binary to merge \citep[e.g.,][]{Antonini&Perets12,NF,Stephan+16,Hoang+18,Rose+19,FragioneFabio19}. Similarly, weak encounters with a passing object may change the eccentricity of the binary and result in a merger \citep[e.g.][]{HamersSamsing19a,Samsing+19,MichaelyPerets19,YoungHamers20}. \citet{Stephan+16} estimates that after a few Gyr, 30 per cent of a GC binary population will have merged through the EKL mechanism. However, if tides work efficiently during these periods of high eccentricity, the binary orbit may instead tighten and circularize \citep[e.g.,][]{NF,Rose+19}. Yet another possibility is that the system undergoes weak eccentricity oscillations \citep[e.g.,][]{Rose+19}.

\section{The Outcome: Relax, Evaporate, or Collide?} \label{sec:outcome}

A competition between several processes determines the dynamical fate of a binary. We consider two examples in Figure~\ref{fig:timescales}. Both of the binaries in the figure are equal mass with $M_\mathrm{bin} = 2 \, M_\odot$. Their semimajor axes differ by a factor of $5$ to illustrate the relationship between these competing timescales and the softness of the binary. These binary parameters ensure that the systems are both long-lived and soft enough to undergo evaporation throughout most of the inner region $r \leq 0.5$~pc of the GC. 

Figure~\ref{fig:timescales} indicates that most soft binaries will evaporate before one of the binary members collides with a passerby. However, within about $10^{-2}$~pc of the SMBH, those systems that approach the boundary $s = 1$, given by Eq.~(\ref{eq:softness}), may undergo an inelastic collision before the system has had time to evaporate. We reserve a detailed examination of this outcome for future study. Additionally, for most of inner parsec of the GC, the evaporation timescale is shorter than the relaxation timescale, implying that the binary will evaporate before it can migrate closer to the SMBH. However, for both of the examples, as $r_\bullet$ approaches $0.5$ pc, the relaxation timescale becomes shorter than the evaporation timescale. Here, a binary migrates inward before it unbinds.




The examples in Figure~\ref{fig:timescales} assume that the binary has a circular orbit about the SMBH. The collision timescale decreases with the eccentricity $e_\bullet$ (Figure~\ref{fig:tcoll_vs_ecc.png}), while the evaporation timescale increases (Figure~\ref{fig:eccentricity}). An eccentric orbit will increase the range of distances from the SMBH over which $t_{\rm coll} < t_{\rm ev}$. Additionally, eccentric orbits have a shorter relaxation timescale than circular orbits of the same semimajor axis \citep{SariFragione19}. Therefore, the relaxation and evaporation timescales will cross closer to the SMBH for eccentric orbits, increasing the parameter space in which the binary migrates before it evaporates.

\begin{figure}
	\includegraphics[width=\columnwidth]{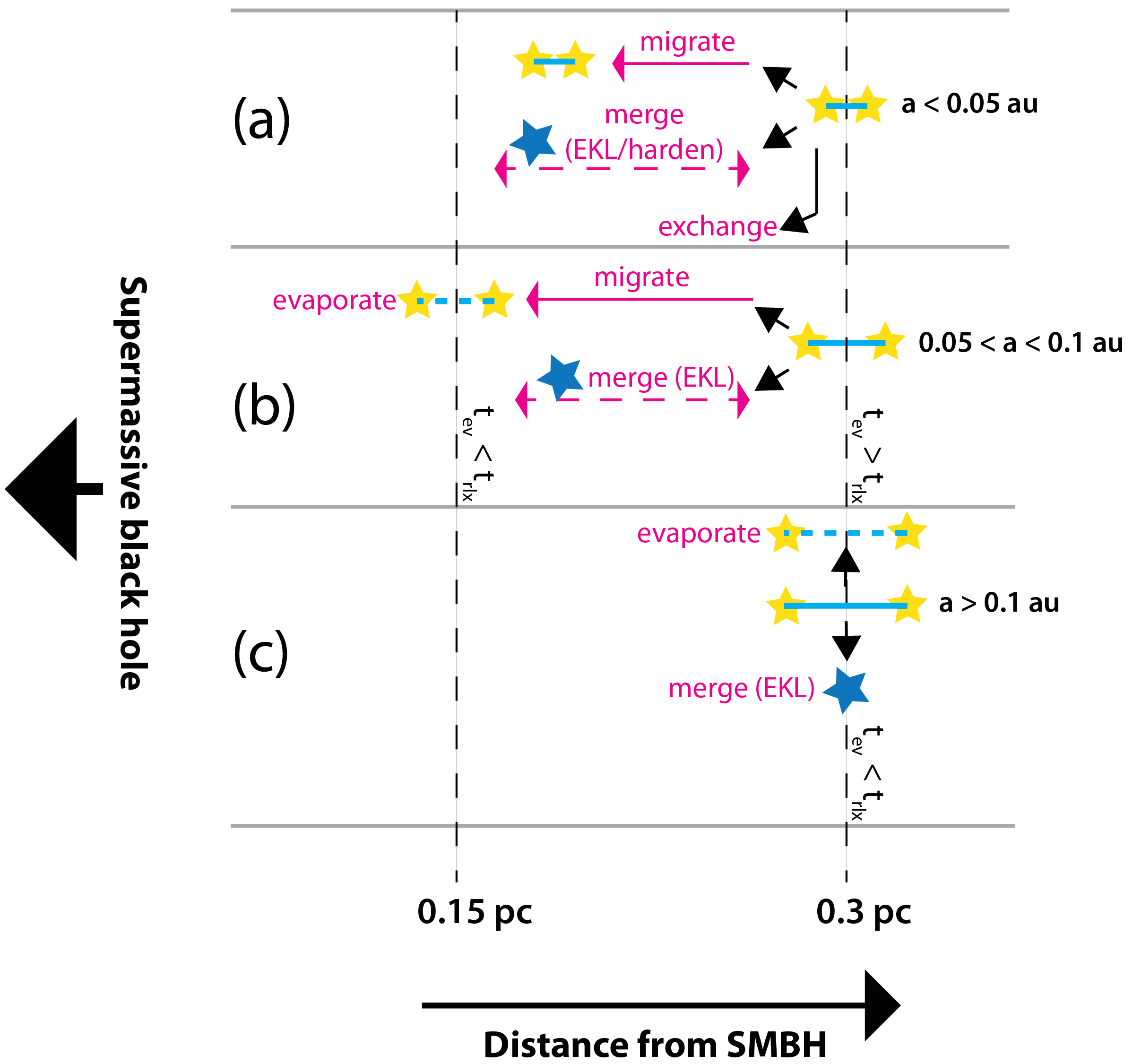}
    \caption{We chart qualitative evolutionary possibilities for a $M_{\rm bin} = 2 \, M_\odot$ binary system at $0.3$~pc from the SMBH depending on its initial semimajor axis, depicted by the blue lines. A dashed blue line indicates a binary that has evaporated. We explore three cases, labeled (a), (b), and (c). Each case corresponds to a row, separated by the light grey solid lines. The shortest timescale at the binary's location determines its fate. The binary migrates when $t_{\rm relax} < t_{\rm ev}$. Note that in scenario (a), the binary is too hard to evaporate. The collision timescale for these systems is too long to alter the qualitative evolution. In all cases, a merger represents a possible outcome. Blue stars symbolize a merger product. Numerical values are approximate in this qualitative illustration.}
    \label{fig:evolution}
\end{figure}

Figure~\ref{fig:evolution} illustrates the qualitative evolutionary trajectories for a binary system depending on its initial semimajor axis $a_{\rm bin}$. This binary resides $0.3$~pc from the SMBH and is an equal mass system with $M_{\rm bin} = 2 \, M_\odot$. We consider three qualitative examples. In the case~(a), the binary with $a_{\rm bin} \lesssim 0.05$~au has a hard configuration. This tight binary will migrate towards the SMBH. The hard binary becomes harder from frequent interactions with nearby objects \citep[e.g.,][]{Heggie75}. The system may harden to the point where it crosses the Roche Limit, or the EKL mechanism may drive the system to merge \citep[e.g.,][]{Stephan+16,Stephan19,Rose+19}. The system may also have an exchange interaction with another star \citep[e.g.][]{Hopman09}. We estimate that this system's exchange interaction timescale is approximately $0.8 \times t_{\rm rlx}$ or $4 \times 10^9$ yr \citep[Eq.~(43) from][]{Hopman09}.

For scenario~(b), the case $0.05 \lesssim a_{\rm bin} \lesssim 0.1$~au, the binary is marginally soft. It evaporates over a timescale longer than the relaxation timescale, allowing the system to move inwards over its lifetime. It will eventually evaporate at a closer distance to the SMBH than the location of its birth. The inward migration brings the binary into a comparatively denser region, where the evaporation timescale is shorter. The coupling of the relaxation and evaporation processes may therefore result in a puzzling binary, one whose age and location suggest that it should have already evaporated. Such a system owes its longevity to originating in a less dense region further from the SMBH.

Finally, case~(c) has $0.1 \, {\rm au} \lesssim a_{\rm bin}$. The evaporation process has a shorter timescale compared to other dynamical processes. The system will unbind due to weak interactions with neighbors. In both soft binary cases, (b) and (c), the system may also merge through the EKL mechanism.

To assess the outcomes in a population of binaries, we generate $70000$ stable systems in the GC using similar parameter distributions to \citet{Stephan+16} (see Appendix~\ref{sec:demographics}). About $99$ per cent of these systems are soft. Only about $3$ per cent of these soft systems experience a collision while the binary is still bound. Using the mass segregation timescale, we estimate that up to $10$ per cent of the binaries drift inwards before they unbind (see Appendix \ref{sec:mass} and Figure~\ref{fig:SMA_migration}). These systems tend to be tighter binaries that reside further from the SMBH. These demographics suggest that mass segregation plays a secondary role to evaporation in shaping the distribution of binaries in the GC, in particular the binary fraction as a function of distance from the SMBH. The vast majority of systems are soft and unbind over shorter timescales. As shown previously using EKL simulations with evaporation in the GC \citep[e.g.,][]{Stephan+16,Hoang+18}, we expect a dearth of binaries closer to the SMBH, a trend that will become more pronounced over time.

\section{Constraining the dark cusp and the binary parameters} \label{sec:constraints}
We build upon the framework presented in \citet{AlexanderPfuhl14} which uses a binary to provide density and relaxation time constraints on the dark cusp in the GC. The evaporation timescale relates key binary and GC properties. Section~\ref{sec:outcome} shows that most binaries in the GC are susceptible to evaporation. Eq.~(\ref{eq:mastereq}) presents a unique opportunity to use a set of known parameters to constrain unknown properties. A confirmed binary can thereby be used to probe the density at the GC. On the other hand, a constraint on the density some distance from the SMBH informs the allowable properties of any hypothetical binary system in that vicinity. 
We explore these relations in the context of an old and a young binary.


\subsection{Galactic Center Stellar Density} \label{sec:density}
The evaporation timescale relates binary properties and the density. The age of the system sets a lower limit on the evaporation timescale: the binary would not be observed if the evaporation timescale were shorter than the system's age. Therefore, a confirmed binary provides an upper limit on the density in its vicinity. Setting $t_{\rm age} = t_{\rm ev}$, we can rearrange Eq.~(\ref{eq:mastereq}) to find $\rho_{\mathrm{max}}$ at the distance $a_\bullet$ from the SMBH:
\begin{eqnarray}
     \rho_{\mathrm{max,ev}} = \frac{\sqrt{3}}{32\sqrt{\pi}G} \frac{\sigma(a_\bullet)}{a_{\rm bin} \ln{\Lambda(a_\bullet)}} \frac{1}{t_{\rm age}} \frac{M_{\rm bin}}{\langle M_\ast \rangle} f(e_\bullet) . \label{eq:rho2}
\end{eqnarray}
  Note that we do not assume a constant Coulomb logarithm. 

Collisions may also unbind the binary system \citep[e.g.,][]{Sigurdsson&Phinney93,Fregeau+04}. A collision can ionize the binary if the incoming star has velocity greater than the critical velocity
\begin{eqnarray} \label{eq:vcrit}
     v_{\rm crit} = \left[ \frac{G(M_{\rm bin}+m_3)}{M_{\rm bin} m_3} \left( \frac{m_1 m_2}{a_{\rm bin}} \right) \right]^{1/2}\ ,
\end{eqnarray}
where $M_{\rm bin} = m_1 + m_2$ and $m_3$ is the mass of the incoming neighbor \citep{Fregeau+04}. We take $m_3$ to be $\langle M_\ast \rangle$. We assume that wherever $\sigma > v_{\rm crit}$, a collision will unbind the binary. Therefore, setting $t_{\rm coll} = t_{\rm age}$ provides another density constraint. We can rearrange the collision timescale Eq.~(\ref{eq:t_coll}) to obtain an upper limit on the density at $a_\bullet$:
\begin{eqnarray} \label{eq:rho_max_coll}
\rho_{\mathrm{max,coll}} &=& \frac{\langle M_\ast \rangle}{\pi \sigma(a_\bullet)} \frac{1}{t_{\rm age}} \nonumber \\ &\times&  \left(f_1(e_\bullet) r_c^2 +f_2(e_\bullet) r_c \frac{2GM_{\rm t}}{\sigma(a_\bullet)^2}\right) ^{-1},
\end{eqnarray}
where $M_{\rm t} = m_1+\langle M_\ast \rangle$. We use the primary mass in this calculation because the more massive star will undergo a collision sooner. Above this density limit, the primary has likely already collided with a passing star, ionizing the system.



The maximum density constraints depend on the average mass of the objects interacting with the binary. Based on work in \citet{AlexanderHopman+09}, \citet{AlexanderPfuhl14} present two equations for $\langle M_\ast \rangle$ as a function of distance from the SMBH. One equation corresponds to a top-heavy initial mass function, while the other derives from a universal initial mass function (IMF). In the former case, $\langle M_\ast \rangle$ is approximately $10 \, M_\odot$ over the range of distances we consider, while the latter equation gives an $\langle M_\ast \rangle$ ranging from $1$ to $2 \, M_\odot$. At $0.15$ pc, where the known binary IRS 16NE resides, $\langle M_\ast \rangle$ equals $1.2 \, M_\odot$ for the universal IMF \citep{AlexanderPfuhl14}.

We present a proof of concept for hypothetical young and old binary systems using these density constraints (Figure \ref{fig:rho}). For these systems, we adopt a circular orbit around the SMBH, such that $f(e)\to 1$. We consider two cases for $\langle M_\ast \rangle$. In the first case, we simply adopt $\langle M_\ast \rangle = 1.2 \, M_\odot$. In the second case, we consider $\langle M_\ast \rangle = 10 \, M_\odot$ to reflect \citet{AlexanderPfuhl14}'s top-heavy IMF. We assume that $10 \, M_\odot$ main-sequence stars dominate the surrounding objects and adopt a radius of approximately $3.5 \, R_\odot$ for these objects to calculate the collision density constraint. However, the top heavy IMF is most relevant within $\sim 0.01$ pc of the SMBH, where an abundance of stellar mass black holes may result in an $\langle M_\ast \rangle$ closer to $10 \, M_\odot$ \citep{Freitag+06,AlexanderHopman+09}. The collisional radius is much smaller for stellar mass black holes. While we restrict ourselves to an $\langle M_\ast \rangle$ that remains constant throughout the GC, this procedure can incorporate an $\langle M_\ast \rangle$ that varies as a function of distance from the SMBH with the form $\langle M_\ast \rangle \propto r_\bullet^{-\beta}$. This change will alter the orbit-averaged result (Section~\ref{sec:ecc_avg}). However, the result should still have the form of a hypergeometric function. We expect that the eccentricity dependence will remain weak.


\subsubsection{The Cusp and Young Binaries}
\begin{figure*}
	\includegraphics[width=0.98\textwidth]{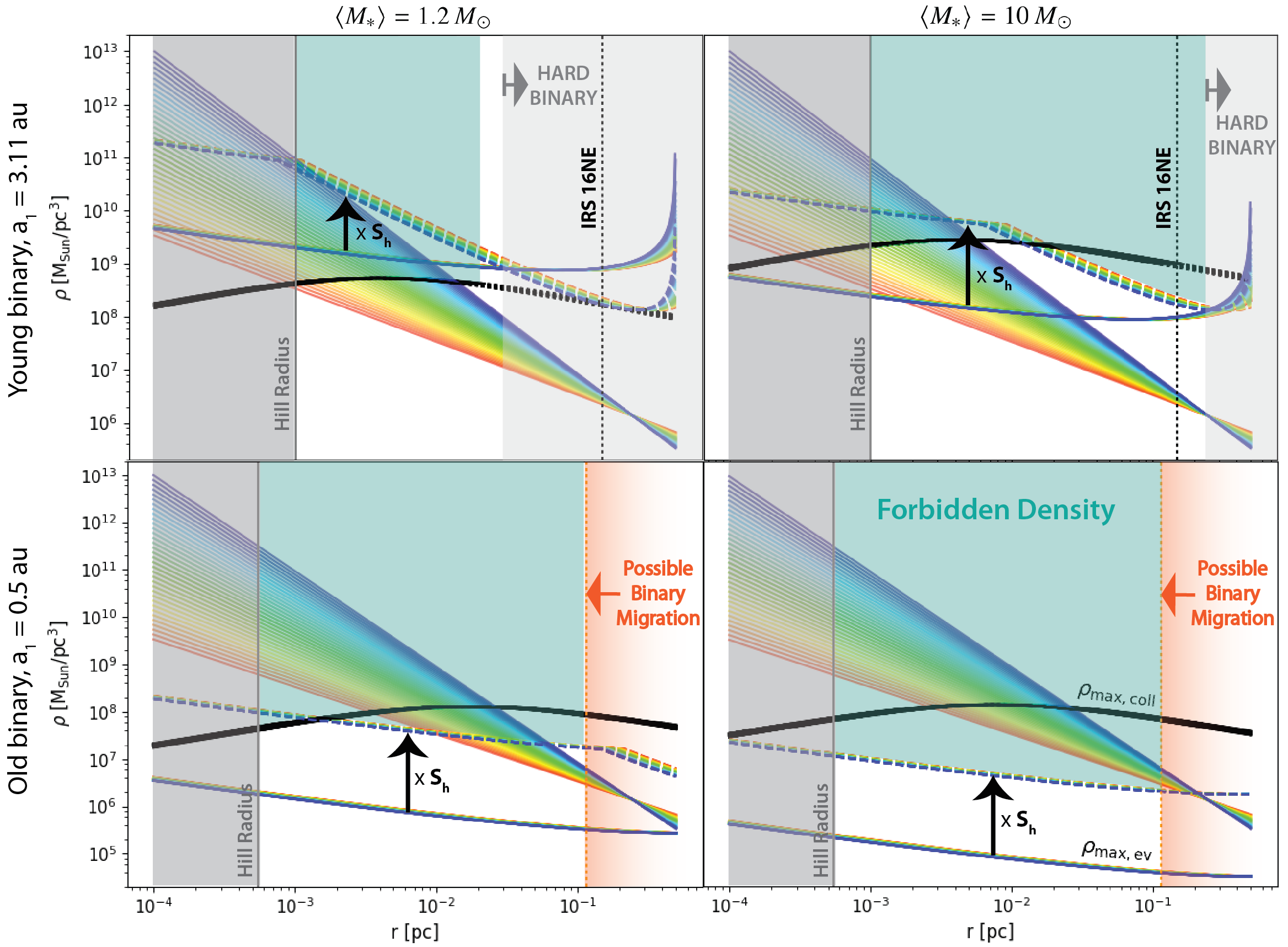}
    \caption{Given a binary, we plot the maximum density for no evaporation as a function of distance from the SMBH with (colorful dashed) and without (colorful solid) the history parameter $S_h$, given by Eq.~(\ref{eq:tevmax}) and (\ref{eq:Sh}). The black curve depicts the maximum density for no collision. The solid portion of the curve denotes the region in which the velocity dispersion is greater than the critical velocity to ionize the binary. The dashed black segment cannot constrain the density. Solid faded lines indicate the density as a function of $r$ assuming a power law like Eq. \ref{eq:density} for $\alpha=1$ (red) to $\alpha=2$ (blue). The left (right) column assumes $\langle M_\ast \rangle = 1.2 \, M_\odot$ ($\langle M_\ast \rangle = 10 \, M_\odot$). See the note at the end of Section~\ref{sec:Evaporation} about the $x$-axes limits, which we extend to extreme values. \textbf{Upper Row: Young Massive Binary} This binary has the parameters of IRS 16NE from \citet{AlexanderPfuhl14}. \textbf{Lower Row: Old Binary} This binary is identical to the system used in the lower panel of Figure \ref{fig:timescales}: a $1$ Gyr old equal mass binary with $1 \, M_\odot$ stars and $0.5$ au separation. Approaching $0.5$ pc, the evaporation timescale exceeds the relaxation timescale, allowing inward migration. These density constraints imply that the detection of an older binary close to the SMBH may suggest a recent dynamical formation scenario.}
    \label{fig:rho}
\end{figure*}

We adopt parameters $M_{\rm bin} = 80 M_\odot$ and $a_{\rm bin} = 3.11$ for the confirmed binary system IRS 16NE \citep{Pfuhl+14,AlexanderPfuhl14}. Like \citet{AlexanderPfuhl14}, we assume an age of $6$ Myr for the system, consistent with observations of the young GC stellar population \citep[e.g.,][]{Paumard+06,Bartko+09,Lu+13}. While IRS 16NE resides at $a_\bullet = 0.15$~pc (see Figure \ref{fig:rho}, black dashed vertical line), we place limits on the density over a range of distances from the SMBH using a IRS 16NE-like binary in the first row of Figure~\ref{fig:rho}.

In conjuction with the known binary parameters, setting $t_{\rm ev} = t_{\rm age}$ gives an upper limit on the density in the binary's vicinity using Eq.~\ref{eq:rho2} (solid multicolored curve; see lower right plot for labels). We scale this density limit by the history parameter to arrive at a conservative maximum density (dashed multicolored curve). The history parameter assumes that the binary has been softening over its lifetime from the tightest possible initial configuration. This density constraint is only valid where the binary is soft enough to evaporate. Gautam et al. (in prep) find that IRS 16NE is a hard binary at $0.15$~pc. For the case $\langle M_\ast \rangle = 1.2 \, M_\odot$, a binary with the parameters $a_{\rm bin}, \, M_{\rm bin}$ of IRS 16NE only becomes soft in the inner $0.01$ pc, where the velocity dispersion is higher. We shade the region where the binary is hard in gray. The left limit of this region depends on the velocity dispersion and therefore $\alpha$. However, to avoid over-cluttering the figure, we simply mark the region as beginning where the binary becomes hard for $\alpha = 2$ (blue).

Similarly, we plot the maximum density constraint from collisions (Eq.~(\ref{eq:rho_max_coll})) in black. The solid segment of the curve coincides with the region in which the neighboring stars have sufficient energy to ionize the system; the velocity dispersion is greater than the critical velocity. The dashed portion indicates the density at which the primary has undergone a collision that may alter the binary configuration while leaving the system bound. The dashed segment therefore cannot constrain the density.

The maximum density constraint is set by either the collision or evaporation timescale, whichever falls lower. We can rule out densities above this threshold. More specifically, we can eliminate certain density power laws of the form Eq.~\ref{eq:density} (faded solid lines) with $\alpha = 1$ (red) to $\alpha = 2$ (blue) that lie above the $\rho_{\rm max}$ curve. We shade the forbidden density region in Figure~\ref{fig:rho}. We note that if the $10 \, M_\odot$ case is dominated by stellar mass black holes as opposed to main sequence stars, the collision timescale becomes very long compared to the evaporation timescale. In this case, the evaporation timescale would set the density constraint in Figure~\ref{fig:rho}.

\subsubsection{The Cusp and Old Binaries}

An older binary places a more stringent constraint on the maximum density than the young system. Given the dearth of confirmed binaries, we use a hypothetical older binary with the same parameters as the lower panel of Figure~\ref{fig:timescales}: $m_1 = m_2 = 1 \, M_\odot$ and $a_{\rm bin} = 0.5$ au. In the orange shaded region, the relaxation timescale exceeds the evaporation timescale, allowing the binary to migrate inward before it has evaporated and placing a caveat on a density constraint derived from the system.\footnote{ Tidal breakup of a triple or quadrupole system can also deliver binary systems from farther out in the GC \citep{FragioneGualandris18,Fragione18}. These systems would similarly provide unreliable density constraints.} However, the left boundary of that region is determined by assuming the binary has evolved from the tightest possible initial configuration. Therefore, it likely overestimates the true region in which a given binary migrates more quickly than it evaporates. We plot $\rho_{\rm max,ev}$ with (dashed dark line) and without (solid dark line) the history parameter. The black dashed curve represents $\rho_{\rm max,coll}$ from Eq.~(\ref{eq:rho_max_coll}). For this long-lived, less massive binary, collisions do not play a significant role in its evolution. Over most of the GC, the binary evaporates before it experiences an inelastic collision.

In Figure~\ref{fig:rho}, we illustrate our interpretation of the plot by shading the region of forbidden density in teal. The maximum density can be compared to the density power laws of the form Eq.~\ref{eq:density} (faded solid lines) with $\alpha = 1$ (red) to $\alpha = 2$ (blue). Within $\sim 0.01$~pc of the SMBH, all density power laws lie above the $\rho_{\rm max}$ for binary survival. We conclude that old binaries cannot survive within $\sim 0.01$~pc of the SMBH. The discovery of a soft, old binary in that region may indicate a dynamical formation mechanism.


\subsection{The Binary Separation} \label{sec:sma}

Similar to the maximum density constraint, we can solve Eq.~(\ref{eq:mastereq}) for $a_{\rm bin}$ to find the maximum separation a binary can have at a certain age by setting $t_{ev} = t_{\rm age}$:
\begin{equation}
\label{eq:amax}
     a_{\mathrm{bin,max}} = \frac{\sqrt{3}}{32\sqrt{\pi}G} \frac{\sigma(r_\bullet)}{\rho(r_\bullet) \ln{\Lambda}}  \frac{1}{t_{\rm age}} \frac{M_{\rm bin}}{\langle M_\ast \rangle}.
\end{equation}
A binary wider than $a_{\mathrm{bin,max}}$ would have already evaporated. We assume that the Coulomb logarithm, which depends on $a_{\rm bin}$, is approximately constant and use a value of 5 based on Figure~\ref{fig:log}. We also assume that $\langle M_\ast \rangle = 1.2 \, M_\odot$ for the remainder of this paper.

In this exercise, we are considering the maximum allowed separation after $t_{\rm age}$ years of evolution. We do not take the history parameter into account because it assumes the tightest possible initial separation for the binary. Evolution from this tight state can only lead to an observed separation shorter then the one presented in Eq.~(\ref{eq:amax}). We also neglect collisions because the collision rate does not depend on the binary semimajor axis. However, a star in these hypothetical binaries may meet the condition $t_{\rm coll} \leq t_{\rm age}$ closer to the SMBH, where the velocity dispersion is high, unbinding the binary. While collisions may act faster than evaporation close to the SMBH, they do not lead to a wider maximum binary separation.


\subsubsection{Young Binary Separation}
\begin{figure}
	\includegraphics[width=\columnwidth]{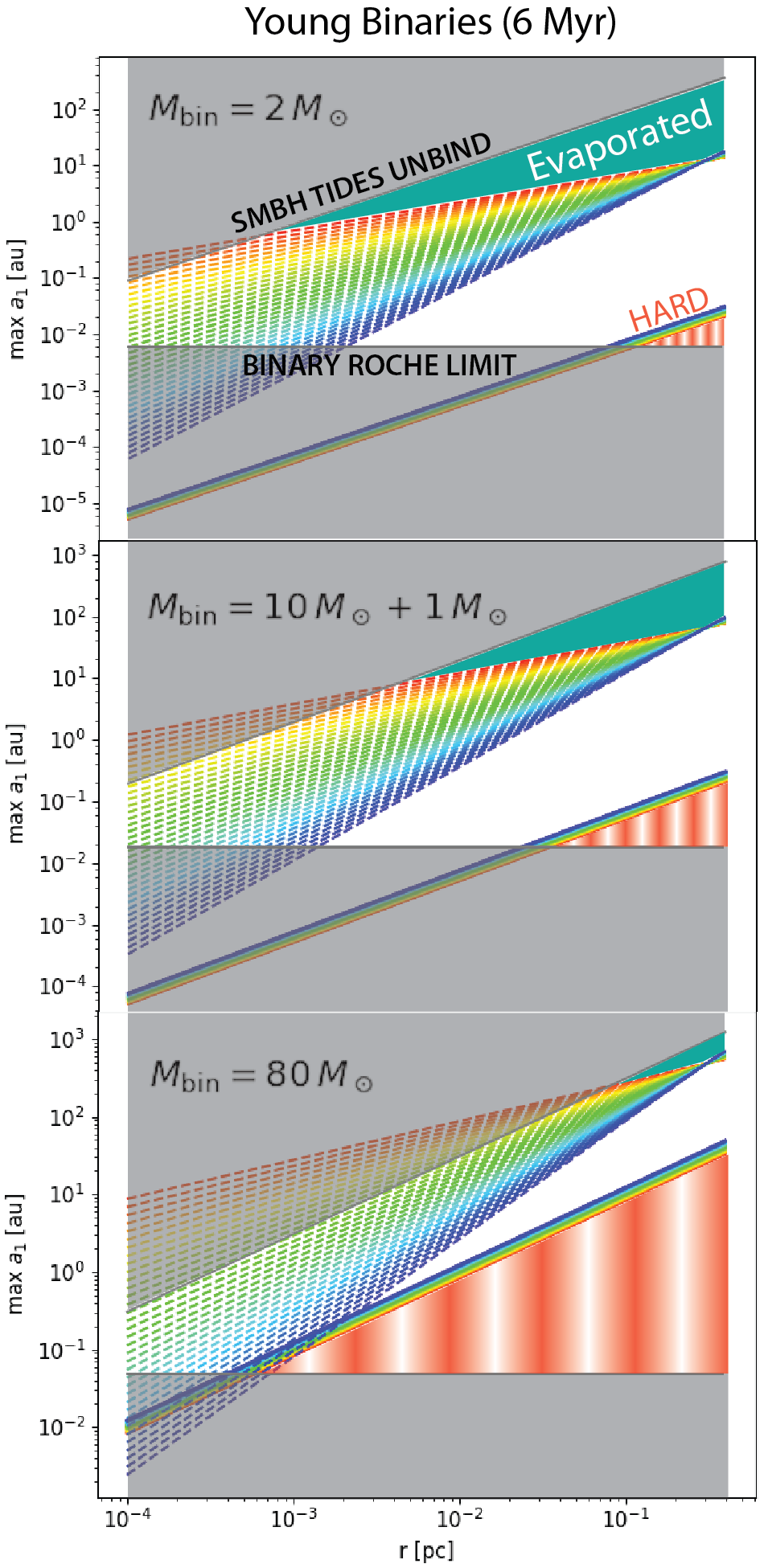}
    \caption{ We consider $M_{\rm bin}=2,11$ and $80$~M$_\odot$ from top to bottom, and estimate the maximum separation after $6.6$~Myr, according to Equation (\ref{eq:amax}), assuming a  constant Coulomb logarithm. This maximum semimajor axis as a function of the binary distance form the SMBH is depicted by the colorful dashed lines, where as in previous figures, we vary $\alpha$ from $1$ (red) to $2$ (blue). Above this line the binary should have been evaporated (teal area). Following the same color convention, we also plot the minimum semimajor axis the binary can have to be considered soft and undergo evaporation. Below this line the binary is hard (shaded with red stripes). We shade in grey the regime at which the inner binary will cross it's own Roche limit (see Eq.~(\ref{eq:RL})) and the regime at which the SMBH  will tidally disrupt and unbind the binary (see Eq.~(\ref{eq:RLSMBH}), where for simplicity we assumed $e_\bullet,e_{\rm bin}\to 0$).  Note the different limits on the $y$-axis due to the different masses. See the note at the end of Section~\ref{sec:Evaporation} about the $x$-axes limits. 
    }
    \label{fig:a1_young}
\end{figure}

In Figure \ref{fig:a1_young}, we consider three young binaries. Two have the same masses as the nominal systems from Section~\ref{sec:density}: one with $m_1 = m_2 = 1 \, M_\odot$ and another with $m_1 = m_2 = 40 \, M_\odot$. We also include a binary with $m_1 = 10 \, M_\odot$ and $m_2 = 1 \, M_\odot$, similar to the parameters in Section~\ref{sec:sstars} where we consider hypothetical binaries in place of S-stars. The dashed colored lines depict the semimajor axis that these young binaries can have as a function of distance from the SMBH for $\alpha = 1$ (red) to $\alpha = 2$ (blue). Above this curve, the binary would have already evaporated within its lifetime. The solid lines, also colored by $\alpha$, represent the minimum semimajor axis the binary can have while remaining susceptible to evaporation. Below this line, the binary is too hard to evaporate. We also plot the limit at which the binary will succumb to tidal forces from the SMBH and become unbound and the limit at which the binary will cross its own Roche limit to undergo mass transfer, labeled the binary Roche limit.


\subsubsection{Old Binary Separation}
Similarly, we plot the maximum $a_1$ (dashed colored lines) for a $1$ Gyr-old binary with $m_1 = m_2 = 1 \, M_\odot$ using to Eq.~(\ref{eq:amax}). Unlike the young binary examples, we omit high-mass stars because they evolve on shorter timescales. As expected, only a narrow range of $a_{\rm bin}$ remains allowable for older systems. Observing an old binary between $0.0025~{\rm pc} \lesssim r_\bullet \lesssim 0.07~{\rm pc}$ may help disentangle the possible power-law index, $\alpha$, of the underlying density. Additionally, observing an old binary within $\sim 0.0025~{\rm pc}$ of the SMBH may suggest a new, perhaps dynamically formed system. 

\begin{figure}
	\includegraphics[width=\columnwidth]{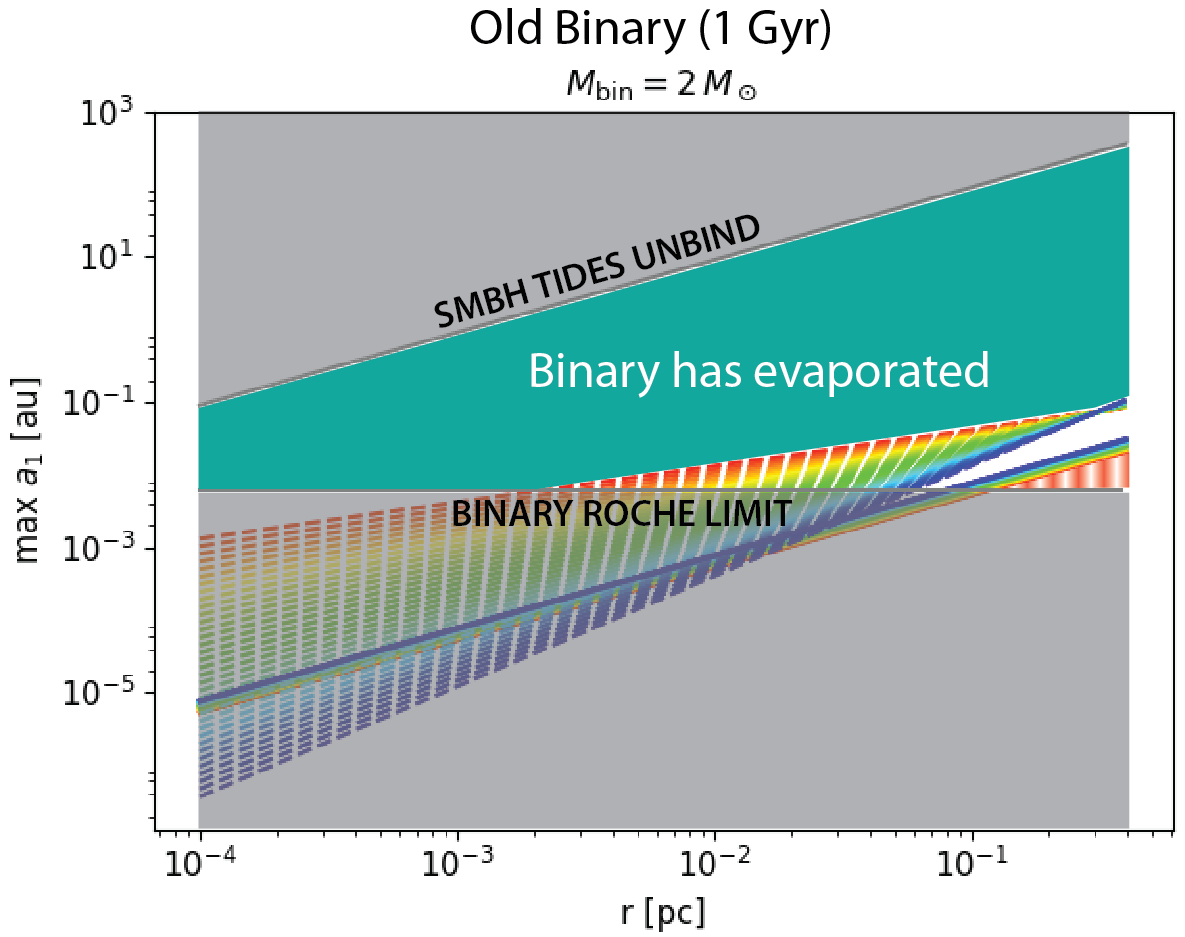}
    \caption{Assuming a constant Coulomb logarithm, the dashed lines indicate the maximum semimajor axis a $1$ Gyr-old binary $2 \, M_\odot$ can have to survive at a certain distance from the SMBH. As in previous figures, we vary $\alpha$ from $1$ (red) to $2$ (blue). We also plot the minimum semimajor axis (solid lines) the binary can have to be considered soft and undergo evaporation. The grey regions represent parameter space excluded either because the binary would be tidally unbound by the SMBH or the stars would undergo mass transfer. The binary is too hard to evaporate in the red striped region. See the note at the end of Section~\ref{sec:Evaporation} about the $x$-axes limits, which we extend to extreme values. }
    \label{fig:a1_old}
\end{figure}






\section{The S-Star cluster} \label{sec:sstars}
While few confirmed binaries exist in the GC, observational evidence indicates their presence. We consider the possibility that observed stars in the GC are embedded in a binary system. Several S-stars have eccentric orbits about the SMBH within $0.05$ pc \citep[e.g.,][]{Ghez+05,Gillessen+09,Gillessen+17}, and observations indicate that many of these stars are young \citep[e.g.,][]{Paumard+06,Lu+09,Bartko+09,Habibi+17}.\footnote{The S-star population also contains some older stars \citep[e.g.][]{Habibi+19}.} Previous studies explore the alternative that these stars, namely S0-2, may in fact represent binary systems and assess their possible orbital configurations \citep[e.g.,][]{Li+17,Chu+18}. The procedure presented here may also constrain the parameter space of the hypothetical binary's orbital configuration. 

\begin{figure*}
	\includegraphics[width=0.9\textwidth]{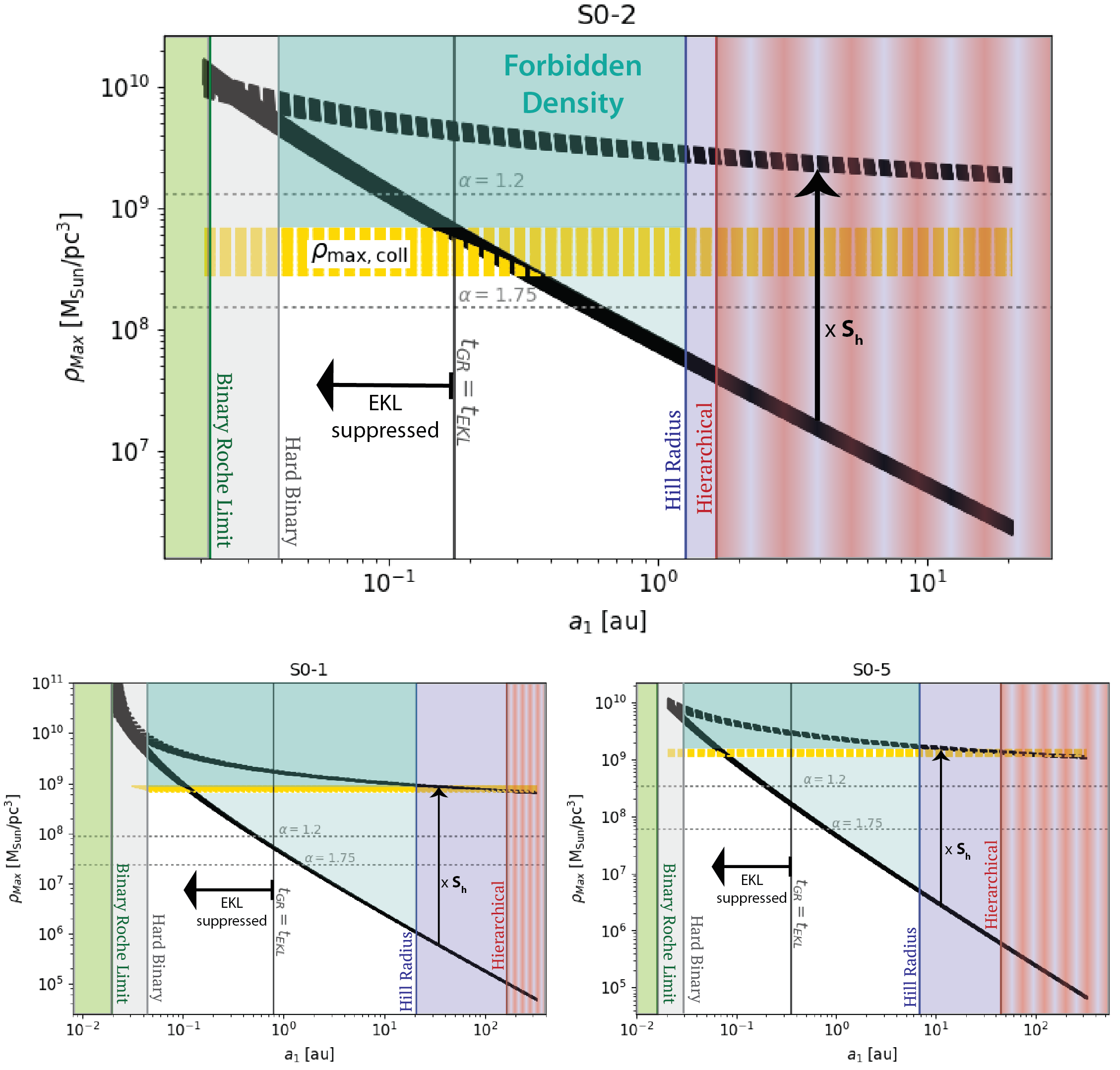}
    \caption{We take the orbital parameters of S0-1, S0-2, and S0-5 and, assuming that these stars in fact represent binaries, constrain the density-semimajor axis parameter space. As a function of binary semimajor axis, we find the maximum density (black solid line) allowed in the region given that the binary has yet to evaporate. We include the history parameter in the dashed black curve. The density at which the primary star has undergone an ionizing collision is given by the horizontal dashed gold line. Criteria for the Hill radius (blue) and dynamical stability (red) place upper limits on the maximum semimajor axis of the hypothetical binary, while the binary's Roche limit (green) sets the lower limit. We also calculated the region in which GR precession suppresses the EKL mechanism. We note that in the S0-1 plot, the $\rho_{\rm max,ev}$ limit does not appear dashed because we plot these curves for a range of $\alpha$.}
    \label{fig:S02}
\end{figure*}

In particular, we consider as a proof of concept three well studied S-stars, S0-1, S0-2, and S0-5 and constrain the allowable semi-major axis and local density (Figure \ref{fig:S02}). A future observational constraint on the density will narrow the range of allowable binary semi-major axis, while a confirmed detection of an S-star binary system will place limits on the density. Radial velocity measurements in particular can provide constraints on the semimajor axes and companion masses of potential S-star binaries (Chu et al. in prep.).

We impose several conditions to ensure that each hypothetical binary is dynamically stable and long-lived. These criteria place upper and lower limits on $a_{\rm bin}$. First, we consider the Hill Radius for a three-body system. This condition requires that the stellar binary's apoapse distance $a_{\rm bin}(1+e_{\rm bin})$ does not exceed the Hill radius:
\begin{equation}\label{eq:RLSMBH}
a_{\rm bin}(1+e_{\rm bin}) < a_\bullet(1-e_\bullet)\left(\frac{m_1}{3M_{\bullet}}\right)^{1/3} \ ,
\label{eq:Hill}
\end{equation}
\citep[e.g.,][]{Naoz+14}. We assume that the hypothetical stellar binary orbit is circular. An eccentric orbit will at most change the maximum $a_{\rm bin}$ by a factor of two. The vertical blue line, labeled accordingly, marks this upper limit in Figure \ref{fig:S02}.

We also impose a hierarchical configuration: the (hypothetical) stellar binary must have a much tighter configuration than the orbit of its center of mass about the SMBH. The SMBH graviationally perturbs the stellar binary and alters its orbital properties. To ensure that these changes are secular and the system is dynamically stable, the parameter $\epsilon$, which represents the pre-factor in the three body Hamiltonian, cannot exceed $0.1$:
\begin{equation}
\epsilon = \frac{a_{\rm bin}}{a_\bullet}\frac{e_\bullet}{1-e_\bullet^2} < 0.1 \ .
\label{eq:epsilon}
\end{equation}
\citep[e.g.,][]{LN11,Naoz16}.  The vertical red line indicates this limit in Figure \ref{fig:S02}. 

Additionally, for a stable binary, no mass transfer or common envelop phase can be occurring. Therefore, the binary cannot cross its own internal Roche Limit. This condition places a lower limit on $a_{\rm bin}$:
\begin{equation}
a_{\rm bin} > R_j\left(\frac{m_j}{m_1+m_2}\right)^{-1/3}, \
\label{eq:RL}
\end{equation}
where $R_j$ denotes the radius of one of the stars. We choose the largest radius between the two stars to reach the most conservative limit, the green vertical line in Figure \ref{fig:S02}.

Radial velocity data for S0-2 (also known as S2) indicate that any hypothetical secondary is limited to about a solar mass \citep{Chu+18}. Furthermore, \citet{Habibi+17} provide values of $6.6$~Myr old and $14 \, M_\odot$ for the age and mass of S0-2. For our hypothetical S0-2 binary, we use a $14 \, M_\odot$ primary and $1 \, M_\odot$ secondary. S0-1 (also known as S1) and S0-5 (also known as S9) have approximate masses $12$ and $8 \, M_\odot$, respectively  \citep[e.g.,][]{Habibi+17}. We use values $0.88$, $0.56$, and $0.64$ for $e_\bullet$ and $16$, $166$, and $51.3$ yr for $P_\bullet$ for S0-2, S0-1, and S0-5, respectively \citep{Gillessen+17}. We assume that the stars formed in the same star formation episode approximately $6$~Myr ago. In our proof of concept, we adopt a secondary of $1 \, M_\odot$ for each hypothetical binary. For less massive stars, the radial velocity effects of a binary may be less apparent, raising the possibility that the star has a more massive dark companion. However, the effects of companion mass on $\rho_{max}$ is relatively weak such that even approaching an equal mass binary, the maximum density will change at most by a factor of $2$.


A stable binary at this proximity to the SMBH will experience EKL perturbations, which tend to increase the binary eccentricity and may lead to merger \citep[e.g.,][]{NF,Naoz+16,Stephan+16,Stephan19}. For example, the  quadrupole timescale for S0-2 falls below $\sim 1 \times 10^{5}$~yr and may be as short as 10~yr for the range of semi-major axis shown in the figure. Thus, it is likely that the EKL mechanism has already driven an S-star binary to merge \citep[e.g.,][]{Stephan+16,Stephan19}. For a binary to remain stable and not undergo EKL oscillations, the  general relativity precession timescale must be shorter than the EKL-quadrapole timescale \citep[e.g.,][]{Ford+00,Naoz+13GR}. Following Eq.~(59) in \citet{Naoz16}, this limit occurs at the black vertical line in Figure \ref{fig:S02}. To the left of this boundary, the EKL mechanism is suppressed.

We calculate the maximum density by using the binary's age as a lower limit on the evaporation (Eq.~(\ref{eq:rho2})) or collision time (Eq.~(\ref{eq:rho_max_coll})) . The solid black curve corresponds to the evaporation constraint. This curve represents the maximum density in the binary's neighborhood as a function of the binary's semi major axis $a_1$. The dashed black curve is the $\rho_{\rm max,ev}$ scaled by the history parameter to give a very conservative upper limit. The true density of the region must lie below this curve for the hypothetical binary to remain bound today. Additionally, the gold curve depicts the density at which the binary is unbound by a physical collision. These hypothetical systems meet the criterion $\sigma (a_{\bullet}) > v_{\rm crit}$ for all values of $\alpha$. The range of $\alpha$ values introduces a spread in the $\rho_{\rm max}$ estimate, with the uppermost limit of $\rho_{\rm max,coll}$ corresponding to  $\alpha = 1$. For comparison, the horizontal dashed grey curves represent the density in the binary's neighborhood as given by the power law Eq.~\ref{eq:density} for two values of $\alpha$.

We note that for younger systems such as these, we expect the true $\rho_{\rm max}$ to lie closer to the solid black curve, $\rho_{\rm max}$ without the history parameter, because the system has had less time to evolve from its initial configuration. We denote this uncertainty by shading this density region in a lighter teal in Figure~\ref{fig:S02}. However, generally, the most conservative maximum density estimate is given by either the collision timescale or the evaporation timescale with the history parameter. For these massive binaries, the collision timescale sets the upper limit on the regional density. The plot can be interpreted in two ways, depending on the observational constraint. For example, if observations indicate a density of $\sim 8 \times 10^8 \, M_\odot/\mathrm{pc}^{3}$, the semimajor axis of an S0-5 binary cannot exceed about $0.1$~au if we assume no evolution from a tighter initial configuration. However, if observations determine that S0-1 is a binary with semimajor axis $0.1$~au, the density must fall below $\sim 10^9 \, M_\odot/\mathrm{pc}^{3}$.


\section{Discussion} \label{sec:conclusion}

Recent observational and theoretical studies suggest the presence of binaries in the GC \citep[e.g.,][]{Ott+99,Rafelski+07,Dong+17a,Dong+17b,Stephan+16,Stephan19,Naoz+18,Gautam+19}. These binaries may have a soft orbital configuration, such that their binding energy is less than the kinetic energy of the neighboring stars. In this dense environment, interactions with neighboring stars can alter the binary's orbital parameters \citep[e.g.,][]{RasioHeggie95,HeggieRasio96}. Over time, these interactions can unbind soft binaries \citep[e.g., see derivation of Eq.~(7.173) in][]{BinneyTremaine}. The timescale for unbinding, or evaporation, depends on the density of the surrounding region. Similarly, a star in a binary may undergo a direct collision with a passing star, which may unbind the binary, and the collision rate depends on the neighborhood density. Therefore, a detection of a soft binary can constrain the underlying density profile in the GC.

We consider the following processes that can affect the binary: (1) unbinding due to interactions with neighboring stars, with the associate evaporation timescale for an arbitrary eccentric orbit (Eq.~(\ref{eq:mastereq})); (2) collision with passing stars (Eq.~(\ref{eq:t_coll})) that may unbind the binary if the velocity dispersion exceeds the critical velocity; (3) the two body relaxation process by which the binary migrates toward the center over a typical timescale (Eq.~(\ref{eq:t_rlx})); and finally, the EKL mechanism that can drive a binary to merge \citep[e.g.,][]{Stephan+16,Stephan19}. Figure \ref{fig:timescales} depicts the relevant timescales. We find that some soft binaries at approximately $0.1$~pc from the SMBH migrate inwards before they unbind. These binaries relocate to a region of higher density, where had they resided over their full lifetime, they would have already evaporated. Additionally, the eccentricity about the SMBH has a marginal effect on the evaporation and collision timescales, used to constrain the GC density.

We derive a density constraint to ensure the binary's survival over its lifetime. Given a binary's age, we estimate the maximum density of the surrounding region, above which the binary would have already evaporated or undergone a direct collision, unbinding the system. We outline this procedure using several proofs of concept. Firstly, we consider the potential of a confirmed wide binary to constrain the GC density. We focus on IRS 16NE, a young wide binary\footnote{\citet{AlexanderPfuhl14} estimate IRS 16NE's age to be consistent with the overall young population at the GC, a few~Myr \citep[e.g.,][]{Paumard+06}.  Furthermore, IRS 16NE has an orbit of $P=224$~days \citep{Pfuhl+14}.}, as a case study. While this binary is hard at its estimated position of $0.15$~pc, we show that an IRS 16NE-like binary can be used to constrain the density closer to the SMBH. We also consider an older low-mass system. Figure~\ref{fig:rho} compares the maximum density allowed by a binary to power law density distributions in the GC.

Additionally, as a binary can constrain environmental properties, environmental properties can inform the maximum orbital separation of a hypothetical binary. Given a power law density distribution for the GC, Figure~\ref{fig:a1_young} and \ref{fig:a1_old} depict the allowable binary separation as a function of distance from the SMBH for a hypothetical system to survive over its lifetime. These figures correspond to a young and old binary, respectively. Figures~\ref{fig:rho} and \ref{fig:a1_old} imply that a low-mass, older ($1 \, {\rm Gyr} \, \leq t_{\rm age}$) binary cannot exist within $0.01$~pc of the SMBH. The detection of such a system may indicate a dynamical formation mechanism. Lastly, we combine the approaches of Figures~\ref{fig:rho} and \ref{fig:a1_young} in Figure~\ref{fig:S02}. The S-star cluster consists of several well-studied stars with constrained orbits about the SMBH \citep[e.g.,][]{Ghez+05,Habibi+19,Gillessen+17}. We consider the possibility that one of these stars is embedded in a soft binary system and constrain the binary separation and density parameter space.

We demonstrate that the density constraining procedure must consider collisions for any system, especially those with a massive star whose larger cross-section increases the collision rate. The collision timescale, not the evaporation timescale, sets the upper limit on the local density, provided that the collision has sufficient energy to ionize the system. Future work may extend this procedure to environments in which a SMBH does not dominate the gravitational potential. In such environments, the velocity dispersion will depend on the density profile. We reserve a more comprehensive integration of collisions and their outcomes into this framework for future study. Future work may also incorporate other processes like tidal capture and clarify their bearing on our ability to derive density constraints from binary systems.

\section*{Acknowledgements}
We dedicate this paper to the late Tal Alexander, who served as an inspiration for the study of dynamics in dense stellar clusters. In particular, SN would like to thank Tal for sponsoring her as a visiting student at the Weizmann Institute. This opportunity helped her, as a young mother, to juggle new responsibilities and pursuing a PhD by reducing her commute. Furthermore, the experience expanded her collaborations and led to new directions. Thank you, Tal!

We thank the referee for useful comments and questions. 
We thank Brad Hansen for asking an important question about collisions, which prompted the collision section. 
SR thanks the Alice Freeman Palmer Fellowship, awarded by Wellesley College, for partial support. SR and SN acknowledge the partial support of NASA grants No. 80NSSC20K0505 and 80NSSC19K0321. SN thanks Howard and Astrid Preston for their generous support.

\appendix

\section{Velocity Dispersion Evolution} \label{sec:demographics}

Our framework does not account for a velocity dispersion that evolves over time. We assume a steady-state. We compute the fractional change in the nuclear star cluster's energy from evaporating binaries, the process that we focus on in this study. However, we stress that this exercise does not provide an accurate picture of the velocity dispersion's time evolution.

\citet{Ciurlo+20} estimate a binary fraction of about $5$ per cent for low-mass stars. Assuming that there are approximately one million stars in the GC, we generate $72000$ dynamically stable stellar binary systems. We use the same birth distributions as \citet{Stephan+16} except our Kroupa IMF has the lower (upper) limit $0.5 \, M_\odot$ ($100 \, M_\odot$) for the primary and secondary masses. We use S0-2's period to set the lower limit on the birth period distribution for the outer orbit \citep{Stephan+16}. Therefore, these binary systems have $a_\bullet \gtrsim 0.003$ or $10^{-2.5}$~pc. For $\alpha = 1.2$, $99$ per cent of the systems are soft. We calculate the evaporation timescale and binding energy for each soft binary. Massive binaries, systems with a $10 \, M_\odot$ or larger primary star, represent $\sim 5$ per cent of the soft systems. All of the soft binaries also meet the criterion to unbind through a collision given by Eq.~(\ref{eq:vcrit}) at their respective distance from the SMBH, however only about $2000$ systems will undergo a direct collision before they evaporate. Additionally, the EKL mechanism should drive some fraction of these systems to merge before they evaporate \citep[e.g.,][]{Stephan+16}. In this calculation, we assume that all soft systems evaporate.

\begin{figure}
    \centering
	\includegraphics[width=0.5\columnwidth]{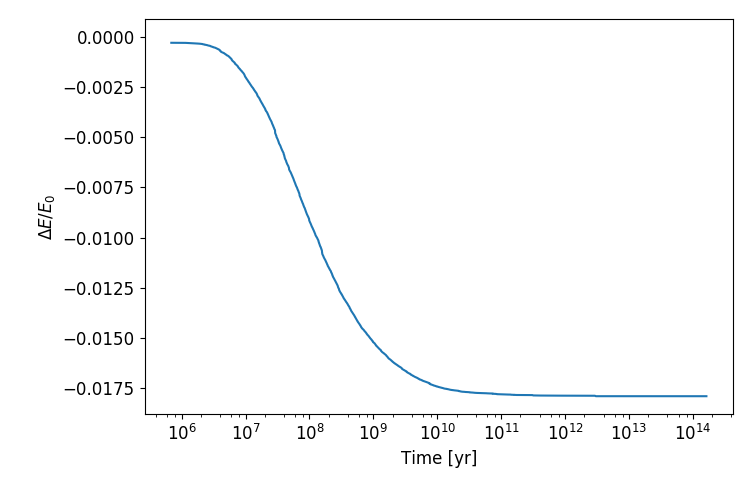}
    \caption{We compute the fractional change in the star cluster's kinetic energy due to evaporating binaries. We compare the binding energy of the evaporated binaries to the total kinetic energy of the star cluster from Eq.~(\ref{eq:E0}).}
    \label{fig:sigma}
\end{figure}

Integrating the following over $10^{-4}$ to $0.5$ pc provides an estimate of the initial kinetic energy for the nuclear star cluster:
\begin{eqnarray} \label{eq:E0}
   E_0 = \frac{1}{2} \int_{r_{\rm inner}}^{r_{\rm outer}} \rho \sigma^2 \times 4\pi r^2 dr,
\end{eqnarray}
where $\rho$ and $\sigma$ are both functions of $r$ given in Section~\ref{sec:maths}. We assume that kinetic energy from the cluster goes into unbinding binaries through interactions with neighbor stars. The kinetic energy of the cluster should therefore decrease with time. We assume that all of the binary's potential energy is released at the evaporation time as opposed to in a gradual process. We use these quantities to arrive at a fractional change in cluster kinetic energy as a function of time (Figure~\ref{fig:sigma}). The small fractional change in energy suggests that a similarly small change in the velocity dispersion, allowing us to assume $\sigma$ is constant over long periods of time such as a low-mass binary's lifetime.


\section{Coulomb Logarithm}

\begin{figure}
    \centering
	\includegraphics[width=0.5\columnwidth]{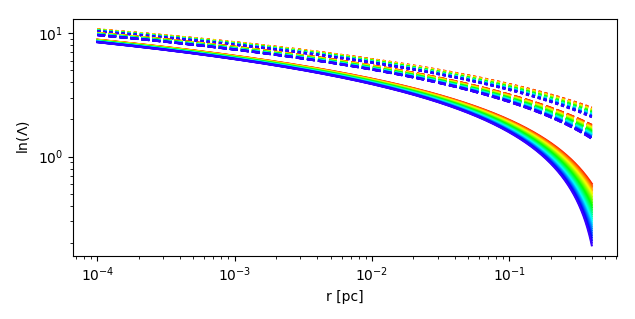}
    \caption{The Coulomb logarithm does not vary substantially within $1$ pc of the Galactic Center. Here we plot the Coulomb logarithm for different values of $\alpha$, ranging from $1$ (red) to $2$ (blue). The dotted curve represents an equal mass $2 \, M_\odot$ binary with separation $0.5$ au. The dashed (solid) curve represent an equal mass  $80 \, M_\odot$ binary with separation $10$ ($3$) au.}
    \label{fig:log}
\end{figure}

We must assume the Coulomb Logarithm is constant in Section~\ref{sec:ecc_avg} to derive the evaporation timescale as a function of eccentricity. This assumption requires that the Coulomb Logarithm varies slowly between the apoapsis and periapsis of an orbit. Figure~\ref{fig:log} shows the Coulomb logarithm for low-mass and massive binaries as a function of distance from the SMBH. Additionally, in Section~\ref{sec:sma}, we assume a constant Coulomb logarithm. We adopt a value of $5$ based on Figure~\ref{fig:log}.

\section{Mass segregation and comparison with globular  clusters}\label{sec:mass}
Unlike globular clusters in which two body relaxation can cause a gradual evaporation of the cluster \citep[e.g.,][]{Gieles+11,Gnedin+14}, nuclear star clusters reside in deeper potential wells. Here we consider a population of $70000$ stable systems (see Appendix~\ref{sec:demographics}).
Figure~\ref{fig:SMA_migration} compares the semimajor axis distributions for the total population of soft binaries (blue) and those with $t_{\rm seg} < t_{\rm ev}$ (orange). The left panel shows $a_{\rm bin}$, and the right, $a_\bullet$. The figure indicates that systems that segregate by mass before unbinding tend to be marginally soft, tighter systems further out from the SMBH, where the velocity dispersion is lower. 

Binary demographics in globular clusters may provide interesting parallels to the GC. For example, 
\citet{Geller+13} and \citet{deGrijs+13} suggest that, contrary to the effects of mass segregation, the binary fraction in the globular cluster NGC 1818 decreases radially towards the center because binary disruption dominates closer to the core. \citet{Cheng+20} discuss a similar competition between mass segregation and binary disruption in the context of the globular cluster M28's X-ray binary population. While N-body simulations are necessary to obtain a comprehensive picture of GC binary demographics over time, based on these calculations, we expect that mass segregation plays a secondary role to evaporation in shaping the distribution of binaries in the GC, in particular the binary fraction as a function of distance from the SMBH. The vast majority of systems are soft and unbind over shorter timescales. We expect a dearth of binaries with decreasing distance to the SMBH, a trend that will become more pronounced over time. Those soft binaries which migrate inwards will also eventually evaporate.

\begin{figure}
    \centering
	\includegraphics[width=0.7\columnwidth]{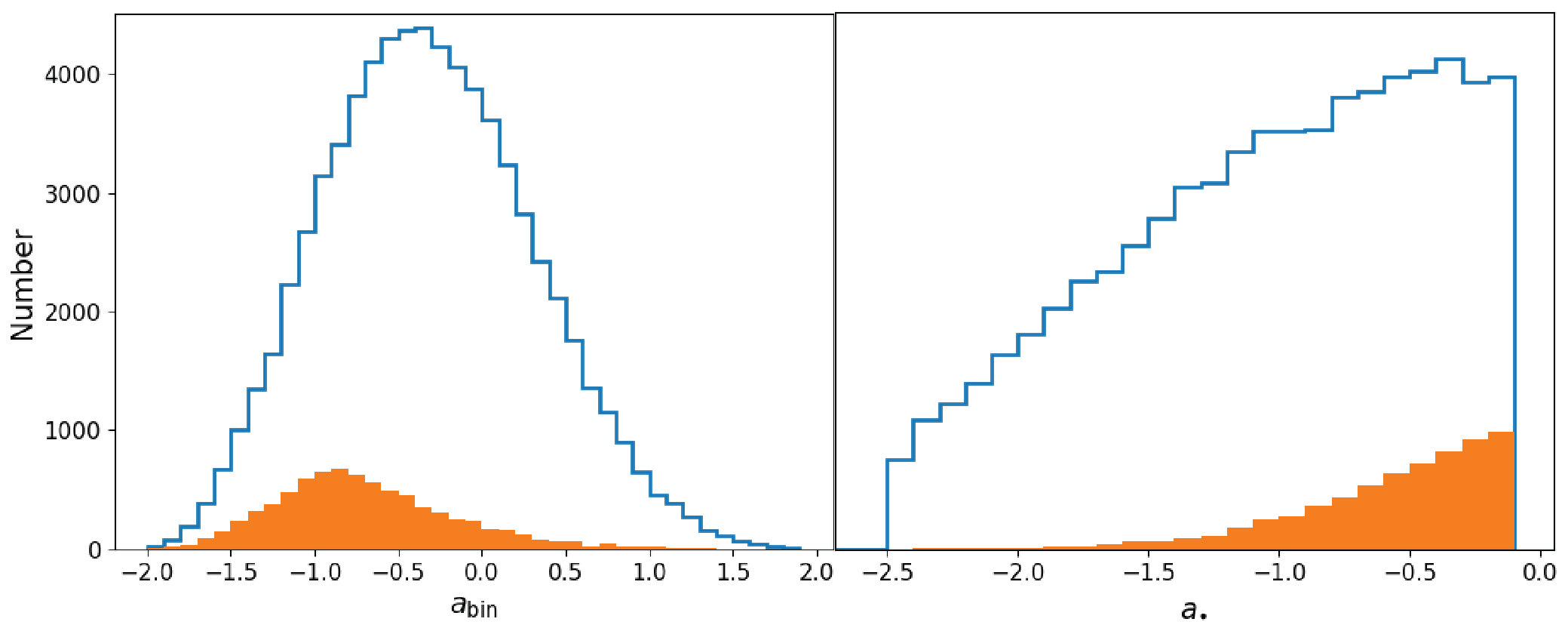}
    \caption{The blue histogram shows the semimajor axes of the inner and outer orbits for approximately $70,000$ soft stellar binaries in the GC using the parameter distributions of \citet{Stephan+16}. The orange filled histogram depicts the semimajor axes for those systems with $t_{\rm seg} < t_{\rm ev}$. The differences between the blue and orange histograms indicate that migrating systems are generally marginally soft, tighter binaries residing further from the SMBH.}
    \label{fig:SMA_migration}
\end{figure}

\bibliographystyle{aasjournal}
\bibliography{EvaporationGC}{}

\begin{thebibliography}{}
\expandafter\ifx\csname natexlab\endcsname\relax\def\natexlab#1{#1}\fi
\providecommand{\url}[1]{\href{#1}{#1}}
\providecommand{\dodoi}[1]{doi:~\href{http://doi.org/#1}{\nolinkurl{#1}}}
\providecommand{\doeprint}[1]{\href{http://ascl.net/#1}{\nolinkurl{http://ascl.net/#1}}}
\providecommand{\doarXiv}[1]{\href{https://arxiv.org/abs/#1}{\nolinkurl{https://arxiv.org/abs/#1}}}

\bibitem[{{Aharon} \& {Perets}(2016)}]{Aharon+16}
{Aharon}, D., \& {Perets}, H.~B. 2016, \apjl, 830, L1,
  \dodoi{10.3847/2041-8205/830/1/L1}

\bibitem[{{Alexander}(1999)}]{Alexander99}
{Alexander}, T. 1999, \apj, 527, 835, \dodoi{10.1086/308129}

\bibitem[{{Alexander} \& {Hopman}(2009)}]{AlexanderHopman+09}
{Alexander}, T., \& {Hopman}, C. 2009, \apj, 697, 1861,
  \dodoi{10.1088/0004-637X/697/2/1861}

\bibitem[{{Alexander} \& {Pfuhl}(2014)}]{AlexanderPfuhl14}
{Alexander}, T., \& {Pfuhl}, O. 2014, \apj, 780, 148,
  \dodoi{10.1088/0004-637X/780/2/148}

\bibitem[{{Ali} {et~al.}(2020){Ali}, {Paul}, {Eckart}, {Parsa}, {Zajacek},
  {Pei{\ss}ker}, {Subroweit}, {Valencia-S.}, {Thomkins}, \&
  {Witzel}}]{Basel+20}
{Ali}, B., {Paul}, D., {Eckart}, A., {et~al.} 2020, \apj, 896, 100,
  \dodoi{10.3847/1538-4357/ab93ae}

\bibitem[{{Antognini}(2015)}]{Antognini15}
{Antognini}, J.~M.~O. 2015, \mnras, 452, 3610, \dodoi{10.1093/mnras/stv1552}

\bibitem[{{Antonini} {et~al.}(2010){Antonini}, {Faber}, {Gualandris}, \&
  {Merritt}}]{Antonini+10}
{Antonini}, F., {Faber}, J., {Gualandris}, A., \& {Merritt}, D. 2010, \apj,
  713, 90, \dodoi{10.1088/0004-637X/713/1/90}

\bibitem[{{Antonini} {et~al.}(2011){Antonini}, {Lombardi}, \&
  {Merritt}}]{Antonini+11}
{Antonini}, F., {Lombardi}, James~C., J., \& {Merritt}, D. 2011, \apj, 731,
  128, \dodoi{10.1088/0004-637X/731/2/128}

\bibitem[{{Antonini} {et~al.}(2014){Antonini}, {Murray}, \&
  {Mikkola}}]{Antonini14}
{Antonini}, F., {Murray}, N., \& {Mikkola}, S. 2014, \apj, 781, 45,
  \dodoi{10.1088/0004-637X/781/1/45}

\bibitem[{{Antonini} \& {Perets}(2012)}]{Antonini&Perets12}
{Antonini}, F., \& {Perets}, H.~B. 2012, \apj, 757, 27,
  \dodoi{10.1088/0004-637X/757/1/27}

\bibitem[{{Bahcall} \& {Wolf}(1976)}]{BahcallWolf76}
{Bahcall}, J.~N., \& {Wolf}, R.~A. 1976, \apj, 209, 214, \dodoi{10.1086/154711}

\bibitem[{{Bar-Or} {et~al.}(2013){Bar-Or}, {Kupi}, \& {Alexander}}]{Bar-Or+13}
{Bar-Or}, B., {Kupi}, G., \& {Alexander}, T. 2013, \apj, 764, 52,
  \dodoi{10.1088/0004-637X/764/1/52}

\bibitem[{{Bartko} {et~al.}(2009){Bartko}, {Martins}, {Fritz}, {Genzel},
  {Levin}, {Perets}, {Paumard}, {Nayakshin}, {Gerhard}, {Alexander},
  {Dodds-Eden}, {Eisenhauer}, {Gillessen}, {Mascetti}, {Ott}, {Perrin},
  {Pfuhl}, {Reid}, {Rouan}, {Sternberg}, \& {Trippe}}]{Bartko+09}
{Bartko}, H., {Martins}, F., {Fritz}, T.~K., {et~al.} 2009, \apj, 697, 1741,
  \dodoi{10.1088/0004-637X/697/2/1741}

\bibitem[{{Bartko} {et~al.}(2010){Bartko}, {Martins}, {Trippe}, {Fritz},
  {Genzel}, {Ott}, {Eisenhauer}, {Gillessen}, {Paumard}, {Alexander},
  {Dodds-Eden}, {Gerhard}, {Levin}, {Mascetti}, {Nayakshin}, {Perets},
  {Perrin}, {Pfuhl}, {Reid}, {Rouan}, {Zilka}, \& {Sternberg}}]{Bartko+10}
{Bartko}, H., {Martins}, F., {Trippe}, S., {et~al.} 2010, \apj, 708, 834,
  \dodoi{10.1088/0004-637X/708/1/834}

\bibitem[{{Baumgardt} {et~al.}(2004){Baumgardt}, {Makino}, \&
  {Ebisuzaki}}]{Baumgardt+04}
{Baumgardt}, H., {Makino}, J., \& {Ebisuzaki}, T. 2004, \apj, 613, 1143,
  \dodoi{10.1086/423299}

\bibitem[{{Binney} \& {Tremaine}(2008)}]{BinneyTremaine}
{Binney}, J., \& {Tremaine}, S. 2008, {Galactic Dynamics: Second Edition}

\bibitem[{{Bonnell} \& {Davies}(1998)}]{BonnellDavies98}
{Bonnell}, I.~A., \& {Davies}, M.~B. 1998, \mnras, 295, 691,
  \dodoi{10.1046/j.1365-8711.1998.01372.x}

\bibitem[{{Buchholz} {et~al.}(2009){Buchholz}, {Sch{\"o}del}, \&
  {Eckart}}]{Buchholz+09}
{Buchholz}, R.~M., {Sch{\"o}del}, R., \& {Eckart}, A. 2009, \aap, 499, 483,
  \dodoi{10.1051/0004-6361/200811497}

\bibitem[{{Cheng} {et~al.}(2018){Cheng}, {Li}, {Xu}, \& {Li}}]{Cheng+18}
{Cheng}, Z., {Li}, Z., {Xu}, X., \& {Li}, X. 2018, \apj, 858, 33,
  \dodoi{10.3847/1538-4357/aaba16}

\bibitem[{{Cheng} {et~al.}(2020){Cheng}, {Mu}, {Li}, {Xu}, {Wang}, \&
  {Li}}]{Cheng+20}
{Cheng}, Z., {Mu}, H., {Li}, Z., {et~al.} 2020, \apj, 892, 16,
  \dodoi{10.3847/1538-4357/ab7933}

\bibitem[{{Chu} {et~al.}(2018){Chu}, {Do}, {Hees}, {Ghez}, {Naoz}, {Witzel},
  {Sakai}, {Chappell}, {Gautam}, {Lu}, \& {Matthews}}]{Chu+18}
{Chu}, D.~S., {Do}, T., {Hees}, A., {et~al.} 2018, \apj, 854, 12,
  \dodoi{10.3847/1538-4357/aaa3eb}

\bibitem[{{Ciurlo} {et~al.}(2020){Ciurlo}, {Campbell}, {Morris}, {Do}, {Ghez},
  {Hees}, {Sitarski}, {Kosmo O'Neil}, {Chu}, {Martinez}, {Naoz}, \&
  {Stephan}}]{Ciurlo+20}
{Ciurlo}, A., {Campbell}, R.~D., {Morris}, M.~R., {et~al.} 2020, \nat, 577,
  337, \dodoi{10.1038/s41586-019-1883-y}

\bibitem[{{Cohn} \& {Kulsrud}(1978)}]{Cohn+78}
{Cohn}, H., \& {Kulsrud}, R.~M. 1978, \apj, 226, 1087, \dodoi{10.1086/156685}

\bibitem[{{de Grijs} {et~al.}(2013){de Grijs}, {Li}, {Zheng}, {Deng}, {Hu},
  {Kouwenhoven}, \& {Wicker}}]{deGrijs+13}
{de Grijs}, R., {Li}, C., {Zheng}, Y., {et~al.} 2013, \apj, 765, 4,
  \dodoi{10.1088/0004-637X/765/1/4}

\bibitem[{{Do} {et~al.}(2009){Do}, {Ghez}, {Morris}, {Lu}, {Matthews}, {Yelda},
  \& {Larkin}}]{Do+09}
{Do}, T., {Ghez}, A.~M., {Morris}, M.~R., {et~al.} 2009, \apj, 703, 1323,
  \dodoi{10.1088/0004-637X/703/2/1323}

\bibitem[{{Do} {et~al.}(2013{\natexlab{a}}){Do}, {Lu}, {Ghez}, {Morris},
  {Yelda}, {Martinez}, {Wright}, \& {Matthews}}]{Do+13a}
{Do}, T., {Lu}, J.~R., {Ghez}, A.~M., {et~al.} 2013{\natexlab{a}}, \apj, 764,
  154, \dodoi{10.1088/0004-637X/764/2/154}

\bibitem[{{Do} {et~al.}(2013{\natexlab{b}}){Do}, {Martinez}, {Yelda}, {Ghez},
  {Bullock}, {Kaplinghat}, {Lu}, {Peter}, \& {Phifer}}]{Do+13b}
{Do}, T., {Martinez}, G.~D., {Yelda}, S., {et~al.} 2013{\natexlab{b}}, \apjl,
  779, L6, \dodoi{10.1088/2041-8205/779/1/L6}

\bibitem[{{Dong} {et~al.}(2017{\natexlab{a}}){Dong}, {Sch{\"o}del}, {Williams},
  {Nogueras-Lara}, {Gallego-Cano}, {Gallego-Calvente}, {Wang}, {Morris}, {Do},
  \& {Ghez}}]{Dong+17a}
{Dong}, H., {Sch{\"o}del}, R., {Williams}, B.~F., {et~al.} 2017{\natexlab{a}},
  \mnras, 470, 3427, \dodoi{10.1093/mnras/stx1436}

\bibitem[{{Dong} {et~al.}(2017{\natexlab{b}}){Dong}, {Sch{\"o}del}, {Williams},
  {Nogueras-Lara}, {Gallego-Cano}, {Gallego-Calvente}, {Wang}, {Rich},
  {Morris}, {Do}, \& {Ghez}}]{Dong+17b}
---. 2017{\natexlab{b}}, \mnras, 471, 3617, \dodoi{10.1093/mnras/stx1836}

\bibitem[{{Feldmeier-Krause} {et~al.}(2015){Feldmeier-Krause}, {Neumayer},
  {Sch{\"o}del}, {Seth}, {Hilker}, {de Zeeuw}, {Kuntschner}, {Walcher},
  {L{\"u}tzgendorf}, \& {Kissler-Patig}}]{FK+15}
{Feldmeier-Krause}, A., {Neumayer}, N., {Sch{\"o}del}, R., {et~al.} 2015, \aap,
  584, A2, \dodoi{10.1051/0004-6361/201526336}

\bibitem[{{Ferrarese} \& {Ford}(2005)}]{FerrareseFord05}
{Ferrarese}, L., \& {Ford}, H. 2005, \ssr, 116, 523,
  \dodoi{10.1007/s11214-005-3947-6}

\bibitem[{{Ford} {et~al.}(2000){Ford}, {Kozinsky}, \& {Rasio}}]{Ford+00}
{Ford}, E.~B., {Kozinsky}, B., \& {Rasio}, F.~A. 2000, \apj, 535, 385,
  \dodoi{10.1086/308815}

\bibitem[{{Fragione}(2018)}]{Fragione18}
{Fragione}, G. 2018, \mnras, 479, 2615, \dodoi{10.1093/mnras/sty1593}

\bibitem[{{Fragione} \& {Antonini}(2019)}]{FragioneFabio19}
{Fragione}, G., \& {Antonini}, F. 2019, \mnras, 488, 728,
  \dodoi{10.1093/mnras/stz1723}

\bibitem[{{Fragione} \& {Gualandris}(2018)}]{FragioneGualandris18}
{Fragione}, G., \& {Gualandris}, A. 2018, \mnras, 475, 4986,
  \dodoi{10.1093/mnras/sty145}

\bibitem[{{Fragione} \& {Sari}(2018)}]{FragioneSari18}
{Fragione}, G., \& {Sari}, R. 2018, \apj, 852, 51,
  \dodoi{10.3847/1538-4357/aaa0d7}

\bibitem[{{Fregeau} {et~al.}(2004){Fregeau}, {Cheung}, {Portegies Zwart}, \&
  {Rasio}}]{Fregeau+04}
{Fregeau}, J.~M., {Cheung}, P., {Portegies Zwart}, S.~F., \& {Rasio}, F.~A.
  2004, \mnras, 352, 1, \dodoi{10.1111/j.1365-2966.2004.07914.x}

\bibitem[{{Freitag} {et~al.}(2006){Freitag}, {Amaro-Seoane}, \&
  {Kalogera}}]{Freitag+06}
{Freitag}, M., {Amaro-Seoane}, P., \& {Kalogera}, V. 2006, \apj, 649, 91,
  \dodoi{10.1086/506193}

\bibitem[{{Gallego-Cano} {et~al.}(2018){Gallego-Cano}, {Sch{\"o}del}, {Dong},
  {Nogueras-Lara}, {Gallego-Calvente}, {Amaro-Seoane}, \&
  {Baumgardt}}]{Gallego+18}
{Gallego-Cano}, E., {Sch{\"o}del}, R., {Dong}, H., {et~al.} 2018, \aap, 609,
  A26, \dodoi{10.1051/0004-6361/201730451}

\bibitem[{{Gallego-Cano} {et~al.}(2020){Gallego-Cano}, {Sch{\"o}del},
  {Nogueras-Lara}, {Dong}, {Shahzamanian}, {Fritz}, {Gallego-Calvente}, \&
  {Neumayer}}]{Gallego+20}
{Gallego-Cano}, E., {Sch{\"o}del}, R., {Nogueras-Lara}, F., {et~al.} 2020,
  \aap, 634, A71, \dodoi{10.1051/0004-6361/201935303}

\bibitem[{{Gautam} {et~al.}(2019){Gautam}, {Do}, {Ghez}, {Morris}, {Martinez},
  {Hosek}, {Lu}, {Sakai}, {Witzel}, {Jia}, {Becklin}, \&
  {Matthews}}]{Gautam+19}
{Gautam}, A.~K., {Do}, T., {Ghez}, A.~M., {et~al.} 2019, \apj, 871, 103,
  \dodoi{10.3847/1538-4357/aaf103}

\bibitem[{{Geller} {et~al.}(2013){Geller}, {de Grijs}, {Li}, \&
  {Hurley}}]{Geller+13}
{Geller}, A.~M., {de Grijs}, R., {Li}, C., \& {Hurley}, J.~R. 2013, \apj, 779,
  30, \dodoi{10.1088/0004-637X/779/1/30}

\bibitem[{{Genzel} {et~al.}(2010){Genzel}, {Eisenhauer}, \&
  {Gillessen}}]{Genzel+10}
{Genzel}, R., {Eisenhauer}, F., \& {Gillessen}, S. 2010, Reviews of Modern
  Physics, 82, 3121, \dodoi{10.1103/RevModPhys.82.3121}

\bibitem[{{Ghez} {et~al.}(2005){Ghez}, {Salim}, {Hornstein}, {Tanner}, {Lu},
  {Morris}, {Becklin}, \& {Duch{\^e}ne}}]{Ghez+05}
{Ghez}, A.~M., {Salim}, S., {Hornstein}, S.~D., {et~al.} 2005, \apj, 620, 744,
  \dodoi{10.1086/427175}

\bibitem[{{Ghez} {et~al.}(2008){Ghez}, {Salim}, {Weinberg}, {Lu}, {Do}, {Dunn},
  {Matthews}, {Morris}, {Yelda}, {Becklin}, {Kremenek}, {Milosavljevic}, \&
  {Naiman}}]{Ghez+08}
{Ghez}, A.~M., {Salim}, S., {Weinberg}, N.~N., {et~al.} 2008, \apj, 689, 1044,
  \dodoi{10.1086/592738}

\bibitem[{{Gieles} {et~al.}(2011){Gieles}, {Heggie}, \& {Zhao}}]{Gieles+11}
{Gieles}, M., {Heggie}, D.~C., \& {Zhao}, H. 2011, \mnras, 413, 2509,
  \dodoi{10.1111/j.1365-2966.2011.18320.x}

\bibitem[{{Gillessen} {et~al.}(2009){Gillessen}, {Eisenhauer}, {Trippe},
  {Alexand er}, {Genzel}, {Martins}, \& {Ott}}]{Gillessen+09}
{Gillessen}, S., {Eisenhauer}, F., {Trippe}, S., {et~al.} 2009, \apj, 692,
  1075, \dodoi{10.1088/0004-637X/692/2/1075}

\bibitem[{{Gillessen} {et~al.}(2017){Gillessen}, {Plewa}, {Eisenhauer}, {Sari},
  {Waisberg}, {Habibi}, {Pfuhl}, {George}, {Dexter}, {von Fellenberg}, {Ott},
  \& {Genzel}}]{Gillessen+17}
{Gillessen}, S., {Plewa}, P.~M., {Eisenhauer}, F., {et~al.} 2017, \apj, 837,
  30, \dodoi{10.3847/1538-4357/aa5c41}

\bibitem[{{Ginsburg} \& {Loeb}(2007)}]{GinsburgLoeb}
{Ginsburg}, I., \& {Loeb}, A. 2007, \mnras, 376, 492,
  \dodoi{10.1111/j.1365-2966.2007.11461.x}

\bibitem[{{Gnedin} {et~al.}(2014){Gnedin}, {Ostriker}, \&
  {Tremaine}}]{Gnedin+14}
{Gnedin}, O.~Y., {Ostriker}, J.~P., \& {Tremaine}, S. 2014, \apj, 785, 71,
  \dodoi{10.1088/0004-637X/785/1/71}

\bibitem[{{Habibi} {et~al.}(2017){Habibi}, {Gillessen}, {Martins},
  {Eisenhauer}, {Plewa}, {Pfuhl}, {George}, {Dexter}, {Waisberg}, {Ott}, {von
  Fellenberg}, {Baub{\"o}ck}, {Jimenez-Rosales}, \& {Genzel}}]{Habibi+17}
{Habibi}, M., {Gillessen}, S., {Martins}, F., {et~al.} 2017, \apj, 847, 120,
  \dodoi{10.3847/1538-4357/aa876f}

\bibitem[{{Habibi} {et~al.}(2019){Habibi}, {Gillessen}, {Pfuhl}, {Eisenhauer},
  {Plewa}, {von Fellenberg}, {Widmann}, {Ott}, {Gao}, {Waisberg},
  {Baub{\"o}ck}, {Jimenez-Rosales}, {Dexter}, {de Zeeuw}, \&
  {Genzel}}]{Habibi+19}
{Habibi}, M., {Gillessen}, S., {Pfuhl}, O., {et~al.} 2019, \apjl, 872, L15,
  \dodoi{10.3847/2041-8213/ab03cf}

\bibitem[{{Hailey} {et~al.}(2018){Hailey}, {Mori}, {Bauer}, {Berkowitz},
  {Hong}, \& {Hord}}]{Hailey+18}
{Hailey}, C.~J., {Mori}, K., {Bauer}, F.~E., {et~al.} 2018, \nat, 556, 70,
  \dodoi{10.1038/nature25029}

\bibitem[{{Hamers} {et~al.}(2018){Hamers}, {Bar-Or}, {Petrovich}, \&
  {Antonini}}]{Hamers+18}
{Hamers}, A.~S., {Bar-Or}, B., {Petrovich}, C., \& {Antonini}, F. 2018, \apj,
  865, 2, \dodoi{10.3847/1538-4357/aadae2}

\bibitem[{{Hamers} \& {Samsing}(2019)}]{HamersSamsing19a}
{Hamers}, A.~S., \& {Samsing}, J. 2019, \mnras, 487, 5630,
  \dodoi{10.1093/mnras/stz1646}

\bibitem[{{Heggie}(1975)}]{Heggie75}
{Heggie}, D.~C. 1975, \mnras, 173, 729, \dodoi{10.1093/mnras/173.3.729}

\bibitem[{{Heggie} \& {Hut}(1993)}]{HeggieHut93}
{Heggie}, D.~C., \& {Hut}, P. 1993, \apjs, 85, 347, \dodoi{10.1086/191768}

\bibitem[{{Heggie} \& {Rasio}(1996)}]{HeggieRasio96}
{Heggie}, D.~C., \& {Rasio}, F.~A. 1996, \mnras, 282, 1064,
  \dodoi{10.1093/mnras/282.3.1064}

\bibitem[{{Heinke} {et~al.}(2008){Heinke}, {Ruiter}, {Muno}, \&
  {Belczynski}}]{Heinke+08}
{Heinke}, C.~O., {Ruiter}, A.~J., {Muno}, M.~P., \& {Belczynski}, K. 2008, in
  American Institute of Physics Conference Series, Vol. 1010, A Population
  Explosion: The Nature \& Evolution of X-ray Binaries in Diverse Environments,
  ed. R.~M. {Bandyopadhyay}, S.~{Wachter}, D.~{Gelino}, \& C.~R. {Gelino},
  136--142, \dodoi{10.1063/1.2945021}

\bibitem[{{Hills}(1975)}]{Hills75}
{Hills}, J.~G. 1975, \aj, 80, 809, \dodoi{10.1086/111815}

\bibitem[{{Hills}(1988)}]{Hills88}
---. 1988, \nat, 331, 687, \dodoi{10.1038/331687a0}

\bibitem[{{Hoang} {et~al.}(2018){Hoang}, {Naoz}, {Kocsis}, {Rasio}, \&
  {Dosopoulou}}]{Hoang+18}
{Hoang}, B.-M., {Naoz}, S., {Kocsis}, B., {Rasio}, F.~A., \& {Dosopoulou}, F.
  2018, \apj, 856, 140, \dodoi{10.3847/1538-4357/aaafce}

\bibitem[{{Hopman}(2009)}]{Hopman09}
{Hopman}, C. 2009, \apj, 700, 1933, \dodoi{10.1088/0004-637X/700/2/1933}

\bibitem[{{Hopman} \& {Alexander}(2006)}]{HopmanAlexander06}
{Hopman}, C., \& {Alexander}, T. 2006, \apj, 645, 1152, \dodoi{10.1086/504400}

\bibitem[{{Keshet} {et~al.}(2009){Keshet}, {Hopman}, \&
  {Alexander}}]{Keshet+09}
{Keshet}, U., {Hopman}, C., \& {Alexander}, T. 2009, \apjl, 698, L64,
  \dodoi{10.1088/0004-637X/698/1/L64}

\bibitem[{{Kocsis} \& {Tremaine}(2011)}]{KocsisTremaine11}
{Kocsis}, B., \& {Tremaine}, S. 2011, \mnras, 412, 187,
  \dodoi{10.1111/j.1365-2966.2010.17897.x}

\bibitem[{{Kormendy}(2004)}]{Kormendy04}
{Kormendy}, J. 2004, in Coevolution of Black Holes and Galaxies, ed. L.~C.
  {Ho}, 1.
\newblock \doarXiv{astro-ph/0306353}

\bibitem[{{Kormendy} \& {Ho}(2013)}]{KormendyHo13}
{Kormendy}, J., \& {Ho}, L.~C. 2013, \araa, 51, 511,
  \dodoi{10.1146/annurev-astro-082708-101811}

\bibitem[{{Kozai}(1962)}]{Kozai}
{Kozai}, Y. 1962, \aj, 67, 591, \dodoi{10.1086/108790}

\bibitem[{{Leigh} {et~al.}(2016){Leigh}, {Antonini}, {Stone}, {Shara}, \&
  {Merritt}}]{Leigh+16}
{Leigh}, N. W.~C., {Antonini}, F., {Stone}, N.~C., {Shara}, M.~M., \&
  {Merritt}, D. 2016, \mnras, 463, 1605, \dodoi{10.1093/mnras/stw2018}

\bibitem[{{Leigh} {et~al.}(2018){Leigh}, {Geller}, {McKernan}, {Ford}, {Mac
  Low}, {Bellovary}, {Haiman}, {Lyra}, {Samsing}, {O'Dowd}, {Kocsis}, \&
  {Endlich}}]{Leigh+18}
{Leigh}, N.~W.~C., {Geller}, A.~M., {McKernan}, B., {et~al.} 2018, \mnras, 474,
  5672, \dodoi{10.1093/mnras/stx3134}

\bibitem[{{Levin} \& {Beloborodov}(2003)}]{Levin+03}
{Levin}, Y., \& {Beloborodov}, A.~M. 2003, \apjl, 590, L33,
  \dodoi{10.1086/376675}

\bibitem[{{Li} {et~al.}(2017){Li}, {Ginsburg}, {Naoz}, \& {Loeb}}]{Li+17}
{Li}, G., {Ginsburg}, I., {Naoz}, S., \& {Loeb}, A. 2017, \apj, 851, 131,
  \dodoi{10.3847/1538-4357/aa9ce7}

\bibitem[{{Lidov}(1962)}]{Lidov}
{Lidov}, M.~L. 1962, planss, 9, 719, \dodoi{10.1016/0032-0633(62)90129-0}

\bibitem[{{Lithwick} \& {Naoz}(2011)}]{LN11}
{Lithwick}, Y., \& {Naoz}, S. 2011, \apj, 742, 94,
  \dodoi{10.1088/0004-637X/742/2/94}

\bibitem[{{Lu} {et~al.}(2013){Lu}, {Do}, {Ghez}, {Morris}, {Yelda}, \&
  {Matthews}}]{Lu+13}
{Lu}, J.~R., {Do}, T., {Ghez}, A.~M., {et~al.} 2013, \apj, 764, 155,
  \dodoi{10.1088/0004-637X/764/2/155}

\bibitem[{{Lu} {et~al.}(2009){Lu}, {Ghez}, {Hornstein}, {Morris}, {Becklin}, \&
  {Matthews}}]{Lu+09}
{Lu}, J.~R., {Ghez}, A.~M., {Hornstein}, S.~D., {et~al.} 2009, \apj, 690, 1463,
  \dodoi{10.1088/0004-637X/690/2/1463}

\bibitem[{{Martins} {et~al.}(2006){Martins}, {Trippe}, {Paumard}, {Ott},
  {Genzel}, {Rauw}, {Eisenhauer}, {Gillessen}, {Maness}, \&
  {Abuter}}]{Martins06}
{Martins}, F., {Trippe}, S., {Paumard}, T., {et~al.} 2006, \apjl, 649, L103,
  \dodoi{10.1086/508328}

\bibitem[{{Mathieu} \& {Latham}(1986)}]{MathieuLatham86}
{Mathieu}, R.~D., \& {Latham}, D.~W. 1986, \aj, 92, 1364,
  \dodoi{10.1086/114269}

\bibitem[{{Merritt}(2006)}]{Merritt06}
{Merritt}, D. 2006, Reports on Progress in Physics, 69, 2513,
  \dodoi{10.1088/0034-4885/69/9/R01}

\bibitem[{{Merritt}(2010)}]{Merritt10}
---. 2010, \apj, 718, 739, \dodoi{10.1088/0004-637X/718/2/739}

\bibitem[{{Michaely} \& {Perets}(2019)}]{MichaelyPerets19}
{Michaely}, E., \& {Perets}, H.~B. 2019, \apjl, 887, L36,
  \dodoi{10.3847/2041-8213/ab5b9b}

\bibitem[{{Miralda-Escud{\'e}} \& {Gould}(2000)}]{MiraldaEGould00}
{Miralda-Escud{\'e}}, J., \& {Gould}, A. 2000, \apj, 545, 847,
  \dodoi{10.1086/317837}

\bibitem[{{Morris}(1993)}]{Morris93}
{Morris}, M. 1993, \apj, 408, 496, \dodoi{10.1086/172607}

\bibitem[{{Muno} {et~al.}(2006){Muno}, {Bauer}, {Bandyopadhyay}, \&
  {Wang}}]{Muno+06}
{Muno}, M.~P., {Bauer}, F.~E., {Bandyopadhyay}, R.~M., \& {Wang}, Q.~D. 2006,
  \apjs, 165, 173, \dodoi{10.1086/504798}

\bibitem[{{Muno} {et~al.}(2005){Muno}, {Pfahl}, {Baganoff}, {Brandt}, {Ghez},
  {Lu}, \& {Morris}}]{Muno+05}
{Muno}, M.~P., {Pfahl}, E., {Baganoff}, F.~K., {et~al.} 2005, \apjl, 622, L113,
  \dodoi{10.1086/429721}

\bibitem[{{Muno} {et~al.}(2009){Muno}, {Bauer}, {Baganoff}, {Band yopadhyay},
  {Bower}, {Brandt}, {Broos}, {Cotera}, {Eikenberry}, {Garmire}, {Hyman},
  {Kassim}, {Lang}, {Lazio}, {Law}, {Mauerhan}, {Morris}, {Nagata},
  {Nishiyama}, {Park}, {Ram{\`\i}rez}, {Stolovy}, {Wijnands}, {Wang}, {Wang},
  \& {Yusef-Zadeh}}]{Muno+09}
{Muno}, M.~P., {Bauer}, F.~E., {Baganoff}, F.~K., {et~al.} 2009, \apjs, 181,
  110, \dodoi{10.1088/0067-0049/181/1/110}

\bibitem[{{Naoz}(2016)}]{Naoz16}
{Naoz}, S. 2016, \araa, 54, 441, \dodoi{10.1146/annurev-astro-081915-023315}

\bibitem[{{Naoz} \& {Fabrycky}(2014)}]{NF}
{Naoz}, S., \& {Fabrycky}, D.~C. 2014, ArXiv e-prints.
\newblock \doarXiv{1405.5223}

\bibitem[{{Naoz} {et~al.}(2016){Naoz}, {Fragos}, {Geller}, {Stephan}, \&
  {Rasio}}]{Naoz+16}
{Naoz}, S., {Fragos}, T., {Geller}, A., {Stephan}, A.~P., \& {Rasio}, F.~A.
  2016, \apjl, 822, L24, \dodoi{10.3847/2041-8205/822/2/L24}

\bibitem[{{Naoz} {et~al.}(2018){Naoz}, {Ghez}, {Hees}, {Do}, {Witzel}, \&
  {Lu}}]{Naoz+18}
{Naoz}, S., {Ghez}, A.~M., {Hees}, A., {et~al.} 2018, \apjl, 853, L24,
  \dodoi{10.3847/2041-8213/aaa6bf}

\bibitem[{{Naoz} {et~al.}(2013){Naoz}, {Kocsis}, {Loeb}, \&
  {Yunes}}]{Naoz+13GR}
{Naoz}, S., {Kocsis}, B., {Loeb}, A., \& {Yunes}, N. 2013, \apj, 773, 187,
  \dodoi{10.1088/0004-637X/773/2/187}

\bibitem[{{Naoz} \& {Silk}(2014)}]{Naoz+14}
{Naoz}, S., \& {Silk}, J. 2014, \apj, 795, 102,
  \dodoi{10.1088/0004-637X/795/2/102}

\bibitem[{{Naoz} {et~al.}(2020){Naoz}, {Will}, {Ramirez-Ruiz}, {Hees}, {Ghez},
  \& {Do}}]{Naoz+20}
{Naoz}, S., {Will}, C.~M., {Ramirez-Ruiz}, E., {et~al.} 2020, \apjl, 888, L8,
  \dodoi{10.3847/2041-8213/ab5e3b}

\bibitem[{{Nogueras-Lara} {et~al.}(2019){Nogueras-Lara}, {Sch{\"o}del},
  {Gallego-Calvente}, {Gallego-Cano}, {Shahzamanian}, {Dong}, {Neumayer},
  {Hilker}, {Najarro}, {Nishiyama}, {Feldmeier-Krause}, {Girard}, \&
  {Cassisi}}]{Nogueras-Lara+19}
{Nogueras-Lara}, F., {Sch{\"o}del}, R., {Gallego-Calvente}, A.~T., {et~al.}
  2019, Nature Astronomy, 4, 377, \dodoi{10.1038/s41550-019-0967-9}

\bibitem[{{Ott} {et~al.}(1999){Ott}, {Eckart}, \& {Genzel}}]{Ott+99}
{Ott}, T., {Eckart}, A., \& {Genzel}, R. 1999, \apj, 523, 248,
  \dodoi{10.1086/307712}

\bibitem[{{Paumard} {et~al.}(2006){Paumard}, {Genzel}, {Martins}, {Nayakshin},
  {Beloborodov}, {Levin}, {Trippe}, {Eisenhauer}, {Ott}, {Gillessen}, {Abuter},
  {Cuadra}, {Alexander}, \& {Sternberg}}]{Paumard+06}
{Paumard}, T., {Genzel}, R., {Martins}, F., {et~al.} 2006, \apj, 643, 1011,
  \dodoi{10.1086/503273}

\bibitem[{{Perets}(2009{\natexlab{a}})}]{Perets09a}
{Perets}, H.~B. 2009{\natexlab{a}}, \apj, 690, 795,
  \dodoi{10.1088/0004-637X/690/1/795}

\bibitem[{{Perets}(2009{\natexlab{b}})}]{Perets09}
---. 2009{\natexlab{b}}, \apj, 698, 1330, \dodoi{10.1088/0004-637X/698/2/1330}

\bibitem[{{Perets} {et~al.}(2007){Perets}, {Hopman}, \&
  {Alexander}}]{Perets+07}
{Perets}, H.~B., {Hopman}, C., \& {Alexander}, T. 2007, \apj, 656, 709,
  \dodoi{10.1086/510377}

\bibitem[{{Pfuhl} {et~al.}(2014){Pfuhl}, {Alexander}, {Gillessen}, {Martins},
  {Genzel}, {Eisenhauer}, {Fritz}, \& {Ott}}]{Pfuhl+14}
{Pfuhl}, O., {Alexander}, T., {Gillessen}, S., {et~al.} 2014, \apj, 782, 101,
  \dodoi{10.1088/0004-637X/782/2/101}

\bibitem[{{Prodan} {et~al.}(2015){Prodan}, {Antonini}, \& {Perets}}]{Prodan+15}
{Prodan}, S., {Antonini}, F., \& {Perets}, H.~B. 2015, \apj, 799, 118,
  \dodoi{10.1088/0004-637X/799/2/118}

\bibitem[{{Rafelski} {et~al.}(2007){Rafelski}, {Ghez}, {Hornstein}, {Lu}, \&
  {Morris}}]{Rafelski+07}
{Rafelski}, M., {Ghez}, A.~M., {Hornstein}, S.~D., {Lu}, J.~R., \& {Morris}, M.
  2007, \apj, 659, 1241, \dodoi{10.1086/512062}

\bibitem[{{Raghavan} {et~al.}(2010){Raghavan}, {McAlister}, {Henry}, {Latham},
  {Marcy}, {Mason}, {Gies}, {White}, \& {ten Brummelaar}}]{Raghavan+10}
{Raghavan}, D., {McAlister}, H.~A., {Henry}, T.~J., {et~al.} 2010, \apjs, 190,
  1, \dodoi{10.1088/0067-0049/190/1/1}

\bibitem[{{Rasio} \& {Heggie}(1995)}]{RasioHeggie95}
{Rasio}, F.~A., \& {Heggie}, D.~C. 1995, \apjl, 445, L133,
  \dodoi{10.1086/187907}

\bibitem[{{Rauch} \& {Tremaine}(1996)}]{RauchTremaine96}
{Rauch}, K.~P., \& {Tremaine}, S. 1996, \na, 1, 149,
  \dodoi{10.1016/S1384-1076(96)00012-7}

\bibitem[{{Rose} {et~al.}(2019){Rose}, {Naoz}, \& {Geller}}]{Rose+19}
{Rose}, S.~C., {Naoz}, S., \& {Geller}, A.~M. 2019, \mnras, 488, 2480,
  \dodoi{10.1093/mnras/stz1846}

\bibitem[{{Samsing} {et~al.}(2019){Samsing}, {Hamers}, \& {Tyles}}]{Samsing+19}
{Samsing}, J., {Hamers}, A.~S., \& {Tyles}, J.~G. 2019, \prd, 100, 043010,
  \dodoi{10.1103/PhysRevD.100.043010}

\bibitem[{{Sari} \& {Fragione}(2019)}]{SariFragione19}
{Sari}, R., \& {Fragione}, G. 2019, \apj, 885, 24,
  \dodoi{10.3847/1538-4357/ab43df}

\bibitem[{{Sch{\"o}del} {et~al.}(2014){Sch{\"o}del}, {Feldmeier}, {Kunneriath},
  {Stolovy}, {Neumayer}, {Amaro-Seoane}, \& {Nishiyama}}]{Schodel+14}
{Sch{\"o}del}, R., {Feldmeier}, A., {Kunneriath}, D., {et~al.} 2014, \aap, 566,
  A47, \dodoi{10.1051/0004-6361/201423481}

\bibitem[{{Sch{\"o}del} {et~al.}(2018){Sch{\"o}del}, {Gallego-Cano}, {Dong},
  {Nogueras-Lara}, {Gallego-Calvente}, {Amaro-Seoane}, \&
  {Baumgardt}}]{Schodel+18}
{Sch{\"o}del}, R., {Gallego-Cano}, E., {Dong}, H., {et~al.} 2018, \aap, 609,
  A27, \dodoi{10.1051/0004-6361/201730452}

\bibitem[{{Sch{\"o}del} {et~al.}(2003){Sch{\"o}del}, {Genzel}, {Ott}, \&
  {Eckart}}]{Schodel+03}
{Sch{\"o}del}, R., {Genzel}, R., {Ott}, T., \& {Eckart}, A. 2003, Astronomische
  Nachrichten Supplement, 324, 535, \dodoi{10.1002/asna.200385048}

\bibitem[{{Sch{\"o}del} {et~al.}(2010){Sch{\"o}del}, {Najarro}, {Muzic}, \&
  {Eckart}}]{Schodel+10}
{Sch{\"o}del}, R., {Najarro}, F., {Muzic}, K., \& {Eckart}, A. 2010, \aap, 511,
  A18, \dodoi{10.1051/0004-6361/200913183}

\bibitem[{{Sch{\"o}del} {et~al.}(2020){Sch{\"o}del}, {Nogueras-Lara},
  {Gallego-Cano}, {Shahzamanian}, {Gallego-Calvente}, \&
  {Gardini}}]{Schodel+20}
{Sch{\"o}del}, R., {Nogueras-Lara}, F., {Gallego-Cano}, E., {et~al.} 2020,
  arXiv e-prints, arXiv:2007.15950.
\newblock \doarXiv{2007.15950}

\bibitem[{{Shapiro} \& {Marchant}(1978)}]{Shapiro+78}
{Shapiro}, S.~L., \& {Marchant}, A.~B. 1978, \apj, 225, 603,
  \dodoi{10.1086/156521}

\bibitem[{{Sigurdsson} \& {Phinney}(1993)}]{Sigurdsson&Phinney93}
{Sigurdsson}, S., \& {Phinney}, E.~S. 1993, \apj, 415, 631,
  \dodoi{10.1086/173190}

\bibitem[{{Spitzer}(1987)}]{Spitzer1987}
{Spitzer}, L. 1987, {Dynamical evolution of globular clusters}

\bibitem[{{Stephan} {et~al.}(2016){Stephan}, {Naoz}, {Ghez}, {Witzel},
  {Sitarski}, {Do}, \& {Kocsis}}]{Stephan+16}
{Stephan}, A.~P., {Naoz}, S., {Ghez}, A.~M., {et~al.} 2016, ArXiv e-prints.
\newblock \doarXiv{1603.02709}

\bibitem[{{Stephan} {et~al.}(2019){Stephan}, {Naoz}, {Ghez}, {Morris},
  {Ciurlo}, {Do}, {Breivik}, {Coughlin}, \& {Rodriguez}}]{Stephan19}
---. 2019, arXiv e-prints, arXiv:1903.00010.
\newblock \doarXiv{1903.00010}

\bibitem[{{Yelda} {et~al.}(2014){Yelda}, {Ghez}, {Lu}, {Do}, {Meyer}, {Morris},
  \& {Matthews}}]{Yelda+14}
{Yelda}, S., {Ghez}, A.~M., {Lu}, J.~R., {et~al.} 2014, \apj, 783, 131,
  \dodoi{10.1088/0004-637X/783/2/131}

\bibitem[{{Young} \& {Hamers}(2020)}]{YoungHamers20}
{Young}, S., \& {Hamers}, A.~S. 2020, arXiv e-prints, arXiv:2006.15023.
\newblock \doarXiv{2006.15023}

\bibitem[{{Yu} \& {Tremaine}(2003)}]{YuTremaine}
{Yu}, Q., \& {Tremaine}, S. 2003, \apj, 599, 1129, \dodoi{10.1086/379546}

\bibitem[{{Zhu} {et~al.}(2018){Zhu}, {Li}, \& {Morris}}]{Zhu+18}
{Zhu}, Z., {Li}, Z., \& {Morris}, M.~R. 2018, \apjs, 235, 26,
  \dodoi{10.3847/1538-4365/aab14f}

\end{thebibliography}



\end{document}